\documentclass[10pt,journal,compsoc]{IEEEtran}

\usepackage[normalem]{ulem}
\usepackage{graphicx}
\usepackage{breakcites}
\usepackage{fontawesome}
\usepackage{float}
\usepackage{caption}
\usepackage{subcaption}
\usepackage{multirow}
\usepackage[ruled,vlined,linesnumbered,noend]{algorithm2e}
\usepackage{ragged2e}
\usepackage{amsmath}
\usepackage{mathtools}
\usepackage{amssymb}
\usepackage{amsthm}
\usepackage{tcolorbox}
\usepackage{booktabs}
\usepackage{tabularx}
\usepackage{color, colortbl}
\usepackage[final]{hyperref}
\hypersetup{
    colorlinks = true, 
    urlcolor = black, 
    linkcolor = blue, 
    citecolor = blue 
}

\renewcommand{\thetable}{\arabic{table}}

\newcommand{\varname}[1]{\mathit{#1}}

\def\tablename{Table}
\def\figurename{Figure}

\begin{document}

\title{Multi-Objective Hardware-Mapping Co-Optimisation for Multi-DNN Workloads on Chiplet-based Accelerators}

\author{Abhijit~Das\IEEEauthorrefmark{1},~\IEEEmembership{Member,~IEEE,}
        Enrico~Russo\IEEEauthorrefmark{1},~\IEEEmembership{Student~Member,~IEEE,}
        and~Maurizio~Palesi,~\IEEEmembership{Senior~Member,~IEEE}
\IEEEcompsocitemizethanks{
\IEEEcompsocthanksitem A. Das is with the Department of Computer Architecture, Universitat Politécnica de Catalunya, Barcelona 08034, Spain.\protect\\
E-mail: abhijit.das@upc.edu
\IEEEcompsocthanksitem E. Russo and M. Palesi are with the Department of Electrical, Electronics and Computer Engineering, University of Catania, Catania 95124, Italy.\protect\\
E-mail: enrico.russo@phd.unict.it; maurizio.palesi@unict.it
\IEEEcompsocthanksitem A. Das and E. Russo are co-first authors and contributed equally. 
}}

\markboth{IEEE Transactions on Computers,~2024}%
{Das \MakeLowercase{\textit{et al.}}: Multi-Objective Hardware-Mapping Co-Optimisation for Multi-DNN Workloads on Chiplet-based Accelerators}

\IEEEtitleabstractindextext{%
\justifying{
\begin{abstract}
The need to efficiently execute different Deep Neural Networks (DNNs) on the same computing platform, coupled with the requirement for easy scalability, makes Multi-Chip Module (MCM)-based accelerators a preferred design choice. Such an accelerator brings together heterogeneous sub-accelerators in the form of chiplets, interconnected by a Network-on-Package (NoP). This paper addresses the challenge of selecting the most suitable sub-accelerators, configuring them, determining their optimal placement in the NoP, and mapping the layers of a predetermined set of DNNs spatially and temporally. The objective is to minimise execution time and energy consumption during parallel execution while also minimising the overall cost, specifically the silicon area, of the accelerator.

This paper presents MOHaM, a framework for multi-objective hardware-mapping co-optimisation for multi-DNN workloads on chiplet-based accelerators. MOHaM exploits a multi-objective evolutionary algorithm that has been specialised for the given problem by incorporating several customised genetic operators. MOHaM is evaluated against state-of-the-art Design Space Exploration (DSE) frameworks on different multi-DNN workload scenarios. The solutions discovered by MOHaM are Pareto optimal compared to those by the state-of-the-art. Specifically, MOHaM-generated accelerator designs can reduce latency by up to $96\%$ and energy by up to $96.12\%$.
\end{abstract}
}

}

\maketitle

\IEEEraisesectionheading{\section{Introduction}}
\IEEEPARstart{D}{eep Neural Network (DNN)} accelerators drive the era of Domain-Specific Architectures (DSAs). From edge~\cite{edge2018tpu} to the cloud~\cite{cloud2021tpu}, they are ubiquitous to enable performance improvement for different applications. To meet the ever-increasing computation demand from emerging applications, bigger DNN accelerators are deployed in data centers. Existing works have either proposed monolithic chip-based designs like Google TPUv4~\cite{cloud2021tpu} or Multi-Chip Module (MCM)-based designs like Simba~\cite{shao2019simba}. The latter is preferred as a scalable and adaptable design paradigm to combine multiple accelerators (as chiplets) for building a big multi-accelerator system. Table~\ref{tab:sota} presents the existing works on multi-accelerator systems~\cite{shao2019simba}\cite{cloud2021tpu}\cite{cerebras2021cs2}\cite{baek2020multi}\cite{choi2020prema}\cite{ghodrati2020planaria}\cite{kwon2021heterogeneous}\cite{liu2022veltair}\cite{kao2022magma}\cite{zeng2022serving}. Here, the second column, \textit{Hetero-Core}, indicates the availability of heterogeneous core or chiplet integration.

\begin{table}
\centering
\caption{Comparison of multi-accelerator systems based on the availability of heterogeneous cores (Hetero-Core), multi-DNN workloads (Mul-DNN), hardware-mapping co-optimisation (HW-MP) and multi-objective exploration (Mul-Obj). A (\faMinus) indicates unavailability of information.}
\label{tab:sota}
\resizebox{\columnwidth}{!}{
\begin{tabular}{@{}ccccc@{}}
\toprule
\textbf{Work} & \textbf{Hetero-Core} & \textbf{Mul-DNN} & \textbf{HW-MP} & \textbf{Mul-Obj} \\
\midrule
Simba~\cite{shao2019simba} & \textcolor{red}{\faClose} & \textcolor{red}{\faClose} & \textcolor{red}{\faClose} & \textcolor{red}{\faClose} \\
TPUv4~\cite{cloud2021tpu} & \textcolor{red}{\faClose} & \faMinus & \textcolor{red}{\faClose} & \textcolor{red}{\faClose} \\
CS-2~\cite{cerebras2021cs2} & \textcolor{red}{\faClose} & \textcolor{red}{\faClose} & \textcolor{red}{\faClose} & \textcolor{red}{\faClose} \\
AI-MT~\cite{baek2020multi} & \textcolor{green}{\faCheck} & \textcolor{green}{\faCheck} & \textcolor{red}{\faClose} & \textcolor{red}{\faClose} \\
PREMA~\cite{choi2020prema} & \textcolor{green}{\faCheck} & \textcolor{green}{\faCheck} & \textcolor{red}{\faClose} & \textcolor{red}{\faClose} \\
Planaria~\cite{ghodrati2020planaria} & \textcolor{red}{\faClose} & \textcolor{green}{\faCheck} & \textcolor{green}{\faCheck} & \textcolor{red}{\faClose} \\
Herald~\cite{kwon2021heterogeneous} & \textcolor{green}{\faCheck} & \textcolor{green}{\faCheck} & \textcolor{green}{\faCheck} & \textcolor{red}{\faClose} \\
VELTAIR~\cite{liu2022veltair} & \textcolor{red}{\faClose} & \textcolor{green}{\faCheck} & \textcolor{red}{\faClose} & \textcolor{red}{\faClose} \\
MAGMA~\cite{kao2022magma} & \textcolor{green}{\faCheck} & \textcolor{green}{\faCheck} & \textcolor{red}{\faClose} & \textcolor{red}{\faClose} \\
H3M~\cite{zeng2022serving} & \textcolor{red}{\faClose} & \textcolor{green}{\faCheck} & \textcolor{green}{\faCheck} & \textcolor{red}{\faClose} \\
\midrule
\textbf{MOHaM} & \textcolor{green}{\faCheck} & \textcolor{green}{\faCheck} & \textcolor{green}{\faCheck} & \textcolor{green}{\faCheck} \\
\bottomrule
\end{tabular}
}
\end{table}

A primary enabler of scalability in accelerators is the support for multi-DNN workloads, where different DNNs are executed together. This becomes a fundamental aspect in multi-accelerator systems as they could house heterogeneous accelerators supporting diverse DNNs. Naturally, the third column, \textit{Mul-DNN}, in Table~\ref{tab:sota} indicates that most existing works support multi-DNN workloads in some form.

Two of the most important factors deciding a DNN accelerator's performance are its hardware configuration and mapping strategy. Their design spaces are extremely large and hence are often optimised independently~\cite{huang2021cosa}\cite{hegde2021mind}\cite{kao2020confuciux}. With multiple heterogeneous accelerators, these design spaces are only becoming larger in multi-accelerator systems. For example, MAGMA~\cite{kao2022magma} reported the design space of mapping alone to be of size $O(1e81)$. However, hardware and mapping are interdependent and, if not optimised together, may lead to sub-par performance when workload changes~\cite{kwon2019understanding}. Furthermore, even if a multi-accelerator system is designed for known workloads, there is extreme heterogeneity in layer shapes and operations~\cite{kwon2021heterogeneous}. Hence, hardware-mapping co-optimisation is critical to tackle the efficiency challenges in multi-accelerator systems and improve their performance. The fourth column, \textit{HW-MP}, in Table~\ref{tab:sota} shows the availability of hardware-mapping co-optimisation in existing multi-accelerator systems.

Designing an accelerator often involves multiple objectives, like latency (delay), throughput, energy, area, temperature, Total Cost of Ownership (TCO), carbon footprint, etc. A convenient way to explore is the aggregation of multiple objectives into one (mono-objective), say, Energy-Delay-Product (EDP). However, a chosen group of objectives might be conflicting with one another; in which case, aggregation does not help. With multi-accelerator systems bringing heterogeneous accelerators together, penalties for aggregation will be higher. Hence, a multi-objective exploration is necessary for identifying multiple solutions to select the most suitable one according to the requirement. Unfortunately, the last column, \textit{Mul-Obj}, in Table~\ref{tab:sota} shows that none of the existing systems has explored the multi-objective space.

\textbf{Contribution:} An ideal multi-accelerator system will have all the columns ticked (\faCheck) in Table~\ref{tab:sota}, i.e., it will have \textit{Hetero-Core} for adaptability, \textit{Mul-DNN} for scalability, \textit{HW-MP} for performance, and \textit{Mul-Obj} for solution variety. However, state-of-the-art falls short in one or more of these requirements due to the unavailability of an orthogonal Design Space Exploration (DSE) framework. This work proposes such a framework for \underline{M}ulti-\underline{O}bjective \underline{Ha}rdware-\underline{M}apping co-optimisation for multi-DNN workloads on chiplet-based accelerators (\textbf{MOHaM})\footnote{https://github.com/Haimrich/moham}. A multi-accelerator system designed using MOHaM will tick all the columns in Table~\ref{tab:sota}. The proposed framework has the following features:
\begin{enumerate}
    \item Given a multi-DNN workload and a library of heterogeneous accelerator templates, MOHaM outputs a Pareto set of multi-accelerator system designs and their associated schedules for the optimal trade-off between objectives. It is to be used at design-time.
    \item MOHaM exploits the NSGA-II~\cite{deb2002fast} multi-objective Genetic Algorithm (GA) and leverages the Timeloop~\cite{parashar2019timeloop} + Accelergy~\cite{wu2019accelergy} cost model for DSE.
    \item MOHaM is for designing multi-accelerator systems where the workload can be known or guessed apriori. The workload, accelerator templates and objectives can be set as per the design requirement. 
\end{enumerate}
\noindent This work makes the following novel contributions:
\begin{enumerate}
    \item It frames the need for adaptability, scalability, performance and solution variety in multi-accelerator systems into an optimisation problem. It develops multiple modules to present MOHaM, the first open-source infrastructure for complete DSE of multi-accelerator systems with known workloads.
    \item It designs several custom GA operators and proposes an optimisation algorithm to enable strategic exploration of the huge design space. This increases the sampling efficiency and makes MOHaM orders of magnitude faster than the exhaustive search.
    \item For the evaluated state-of-the-art, MOHaM-generated system designs are capable of reducing latency by $18.40\%$, $13\%$ and $96\%$ and energy by $36.87\%$, $96.12\%$ and $93.76\%$ compared to CoSA~\cite{huang2021cosa}, DiGamma~\cite{kao2022digamma} and Herald~\cite{kwon2021heterogeneous}, respectively.
\end{enumerate}
\section{Background and Motivation}

\begin{figure*}[t]
     \centering
     \begin{subfigure}[b]{0.31\textwidth}
         \centering
         \includegraphics[width=\textwidth]{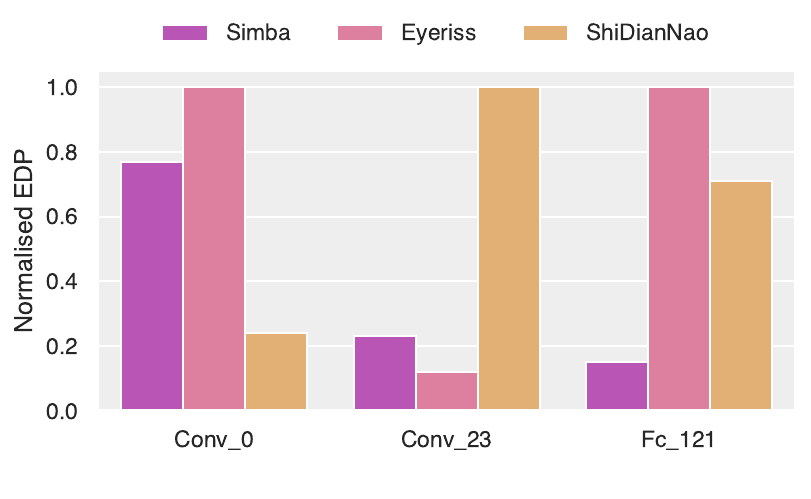}
         \caption{}
         \label{gra:motive1}
     \end{subfigure}
     \hfill
     \begin{subfigure}[b]{0.31\textwidth}
         \centering
         \includegraphics[width=\textwidth]{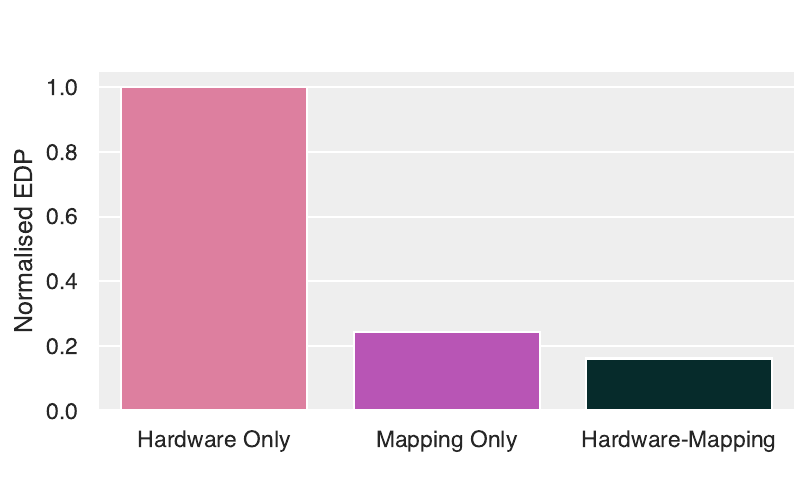}
         \caption{}
         \label{gra:motive2}
     \end{subfigure}
     \hfill
     \begin{subfigure}[b]{0.355\textwidth}
         \centering
         \includegraphics[width=\textwidth]{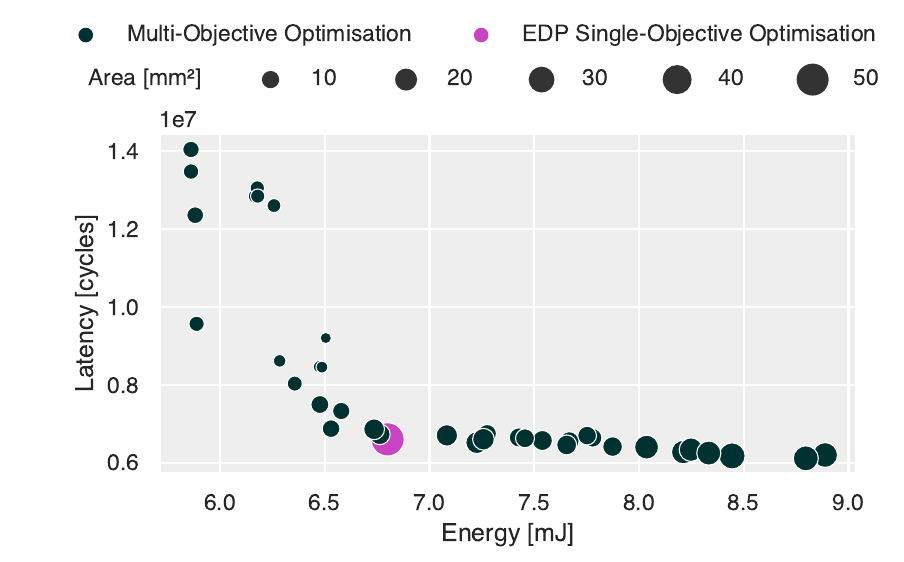}
         \caption{}
         \label{gra:motive3}
     \end{subfigure}
     \caption{Need for heterogeneity, co-optimisation and multi-objective exploration.}
     \label{gra:motive}
\end{figure*}

\subsection{Chiplet-based Accelerators}
To ensure scalability, Simba~\cite{shao2019simba} proposed a Multi-Chip-Module (MCM) based multi-accelerator system, where each chiplet is a DNN accelerator (called \textit{sub-accelerator} in the bigger context) connected by a Network-on-Package (NoP). Each Sub-Accelerator (SA) has an array of Processing Elements (PEs) and a shared Global Buffer (GB)~\cite{chen2016eyeriss}. Each PE houses one or more Multiply-Accumulate (MAC) units to compute partial sums and a Local Buffer (LB) to store them. GB collects weights and activations from memory (HBM/DRAM) through the NoP and distributes them to the LBs through the Network-on-Chip (NoC). Similarly, outputs are written from LB to GB and then to memory through NoC and NoP, respectively. However, unlike Simba, which employs homogeneous chiplets, the proposed MOHaM framework supports heterogeneous chiplets to ensure adaptability with diverse DNNs. Existing literature has many promising accelerators that could be used as SA chiplets in MOHaM for DSE~\cite{qin2020sigma}\cite{shao2019simba}\cite{chen2019eyeriss}\cite{nvdia2018nvdla}\cite{parashar2017scnn}\cite{han2016eie}\cite{du2015shidiannao}.

\subsection{Multi-DNN Workloads}
To maintain competitive performance, emerging applications started employing DNNs for performing tasks like image classification, image segmentation, object detection, etc. For example, an AR/VR application could employ diverse DNNs together for different tasks, thus forming a multi-DNN workload. This diversity of DNNs leads to extreme heterogeneity in layer shapes and operations. Hence, different layers of DNNs could have specific dataflow preferences. However, most of the SAs are usually optimised for some specific layers with a fixed dataflow~\cite{kwon2021heterogeneous}. Figure~\ref{gra:motive1} shows how the normalised EDP varies when the same DNN is run on different SAs. Here, three dominant layers, Convolution \textit{CONV\_0} and \textit{CONV\_33}, and Fully Connected \textit{FC\_121} from the ResNet50 DNN model~\cite{he2016deep} are considered. These layers are individually run on row-stationary (Eyeriss~\cite{chen2016eyeriss}), weight-stationary (Simba~\cite{shao2019simba}) and output-stationary (ShiDianNao~\cite{du2015shidiannao}) SAs. While similar observations are also available in the literature~\cite{kwon2021heterogeneous}\cite{yang2020interstellar}\cite{parashar2019timeloop}\cite{kwon2019understanding}, but the objective here is to advocate for flexible accelerators. MOHaM offers dataflow flexibility with heterogeneous SAs.

\subsection{Hardware-Mapping Co-Optimisation}
A typical hardware-mapping co-optimisation framework takes as input, a target DNN, an optimisation objective (e.g. latency) and the resource constraints (e.g. area). It returns as output a hardware configuration with the optimal instances of resources and an optimal mapping strategy to run the DNN on the accelerator. This co-optimisation can be used at \textit{design-time}. Figure~\ref{gra:motive2} shows how the normalised EDP varies when the same ResNet50 DNN model~\cite{he2016deep} is run with hardware-only, mapping-only and hardware-mapping co-optimisation. Hardware-only optimisation has lower energy due to optimal hardware resources but higher latency (delay) due to fixed mapping. On the contrary, mapping-only optimisation has a lower delay due to optimal mapping strategy but higher energy due to fixed hardware. Co-optimisation gets the best of hardware and mapping, and has the lowest normalised EDP. However, the search space is extremely large as it is the cross-product of hardware and mapping. For example, DiGamma~\cite{kao2022digamma} recently reported a design space as large as $O(10^{36})$, and that is only for a single accelerator. MOHaM offers a DSE which is orders of magnitude faster than the standard exhaustive search.

\subsection{Multi-Objective Exploration}
The convenient way of aggregating multiple optimisation objectives into one (mono-objective) invites penalty when objectives conflict. For example, due to the chosen memory hierarchy, if the system bandwidth is insufficient to feed enough data into the compute (MAC) units, they remain underutilised. In such a scenario, two mapping strategies may output identical latency but varying hardware resource utilisation, hence different energy and area. Similarly, it is possible to either access an enlarged on-chip buffer or the next-level buffer (memory) within the same energy budget. In such a scenario, even two hardware-mapping co-optimisations may output identical energy but different latency and area. These scenarios can be expressed as a set of non-dominating design points (multi-objective). Figure~\ref{gra:motive3} shows the DSE when the same ResNet50 DNN model~\cite{he2016deep} is run for mono-objective exploration with EDP and multi-objective exploration with latency, energy and area. While mono-objective offers an aggregated solution, the Pareto set of distinct multi-objective solutions allows simultaneously exploring accelerator designs with varying requirements. MOHaM offers multi-objective exploration-enabled DSE.
\section{Problem Formulation}

\begin{figure*}
\centering
\includegraphics[width=\textwidth]{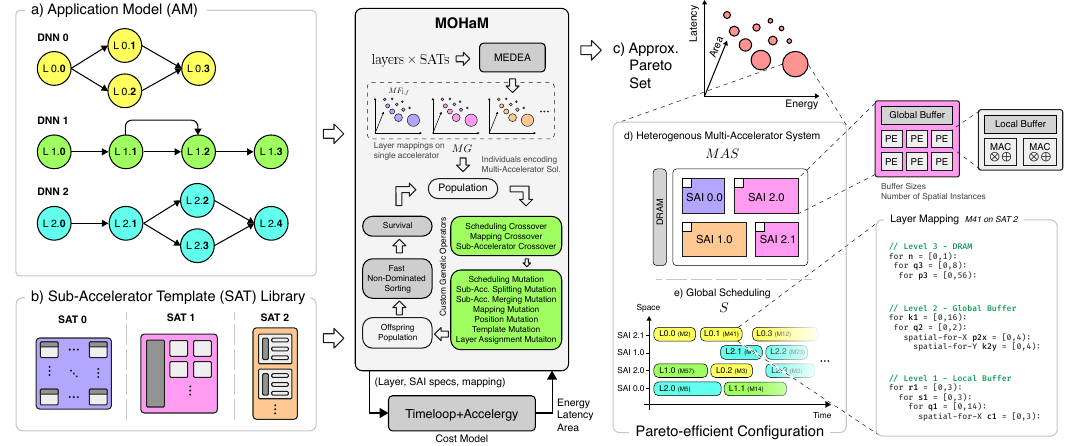}
\caption{Inputs, output and internal organisation of the proposed MOHaM framework.}
\label{fig:moham}
\end{figure*}

Given a set of DNNs to be executed in parallel, and a library of heterogeneous accelerator templates, MOHaM aims to identify a Pareto set of system configurations with associated scheduling for the optimal trade-off between latency, energy and area\footnote{Only for the sake of evaluation, latency, energy and area objectives are
chosen in the problem formulation. The DNNs, accelerator templates and objectives can be modified in MOHaM for specific DSE.}. Figure~\ref{fig:moham} shows MOHaM's internal organisation with inputs and output and are explained here.

\subsection{MOHaM Inputs}
The set of DNNs to be executed in parallel is called an Application Model, \emph{AM}. While layers of different DNNs can be executed in any order, layers within the same DNN must be executed as per their dependencies. Figure~\ref{fig:moham}(a) shows an \emph{AM} with three DNNs, where arrows indicate the order of layer dependencies. For example, in \emph{DNN0}, layer \emph{L0.3} must be executed only after the completion of \emph{L0.1} and \emph{L0.2}.

The library of heterogeneous accelerator templates is a collection of parameterised, reconfigurable DNN accelerators used as chiplets. Each is referred to as a Sub-Accelerator Template, \emph{SAT}, but when its parameters are set, it becomes a Sub-Accelerator Instance, \emph{SAI}. Figure~\ref{fig:moham}(b) shows a library of three \emph{SAT}s. Eyeriss~\cite{chen2016eyeriss}, Simba~\cite{shao2019simba} and ShiDianNao~\cite{du2015shidiannao} \emph{SAT}s are used to evaluate MOHaM (refer Table~\ref{tab:mohamconfig}). For example, in the Eyeriss \emph{SAT}, if the number of PEs, shared buffer size and PE scratchpad size are set to something within their maximum bound, it becomes an Eyeriss \emph{SAI}.

\subsection{MOHaM Output}
A Pareto set of solutions is identified with optimised objectives, latency, energy and area. Figure~\ref{fig:moham}(c) shows a Pareto set, where each solution is a pair of a Multi-Accelerator System, \emph{MAS}, and its associated Schedule, \emph{S}. A \emph{MAS} defines the hardware configuration, like the set of \emph{SAI}s, Memory Interfaces, \emph{MI}s, and their placement in the NoP. Figure~\ref{fig:moham}(d) shows a \emph{MAS} with four heterogenous \emph{SAI}s and an external memory (e.g., DRAM) interconnected by a 2D Mesh NoP.

A schedule \emph{S} defines the mapping strategy, like where each layer of every DNN is mapped in the chiplets, and the order in which they will be executed. Figure~\ref{fig:moham}(e) shows a schedule, where multiple layers are mapped onto the same chiplet (or \emph{SAI}). For example, \emph{L2.0} and \emph{L1.1} are mapped onto \emph{SAI0.0}. In such a case, they are executed sequentially. Each layer gets the mapping information in a loop-nest form.

\subsection{MOHaM Working}
MOHaM takes each layer of the \emph{AM} and maps them to each of the \emph{SAT}s from the input library. It identifies the Pareto set of mappings for each layer across all the \emph{SAT}s with respect to the chosen optimisation objectives. MOHaM then employs a custom GA to conduct multi-objective optimisation with respect to the distinct objectives, latency, energy, and area. This optimisation fine-tunes the (a) selection of Pareto mappings for each layer identified initially, (b) layer scheduling strategies, (c) instantiations of \emph{SAT}s, (d) allocation of hardware resources to each \emph{SAI}, (e) assignment of layers to \emph{SAI}s, and (f) placement of \emph{SAI} tiles in the NoP.  

\subsection{MOHaM Problem Formulation}
Based on the above information, the problem can be formulated as follows: MOHaM aims to minimise latency, energy, and area while ensuring they remain distinct from one another, thereby adopting a multi-objective optimisation.
\begin{eqnarray*}    
    \min \mathit{Latency}(\mathit{AM},\mathit{MAS},\mathit{S}) \\
    \min \mathit{Energy}(\mathit{AM},\mathit{MAS},\mathit{S}) \\
    \min \mathit{Area}(\mathit{MAS})
\end{eqnarray*}
While latency and energy are influenced by factors such as the application model, the configuration of the multi-accelerator system, and the schedule, area is solely determined by the configuration of the multi-accelerator system.

Schedule \emph{S} provides a set of pairs \emph{(L, SAI)}, indicating which layer is mapped to which SA instance along with the order of their execution. A layer can not be mapped into more than one SA instance. It can be represented by:
\[ \forall \ t \ \nexists \ (L_i, SAI_i) \in S \land (L_j, SAI_j) \in S \]
such that
\[ L_i = L_j \land SAI_i \ne SAI_j \]

\subsection{MOHaM Search Space}
Let \emph{Num. Layers} and \emph{Num. DNNs} be the number of layers and DNNs, respectively, that constitute an application model. Let \emph{Max. Chiplets} be the maximum chiplets that can be used in a multi-accelerator system. Let \emph{Num. SATs} be the number of templates available in the \emph{SAT} library. Let \emph{Num. Params} and \emph{Num. Values} be the number of free parameters and the number of values each parameter can take, respectively, of a \emph{SAT}. For the sake of simplification, it is assumed that \emph{Num. Params} and \emph{Num. Values} are the same for all the \emph{SAT}s\footnote{It compacts term~\eqref{eqt:param_space} and avoids its expansion for each \emph{SAT}.}. A lower bound for the search space size is:
\begin{gather}
\varname{Search \ Space \ Size} \ \geq \left( 7! \times \prod_{d=1}^7 m^{ \lvert F_d \rvert} \right)^{\varname{Num. \ Layers}}  \label{eqt:nmappings} \\
\times \sum_{\varname{Chiplets}=1}^{\varname{Max. \ Chiplets}} \Big( \varname{Num. \ SATs}^{\varname{Chiplets}} \label{eqt:chiplet_template} \\
\times \varname{\ Num. \ Values}^{\varname{Num. \ Params \ } \times \varname{ \ Chiplets}} \label{eqt:param_space} \\
\times {\varname{\ Chiplets}}! \label{eqt:chiplet_placement} \\
\times \varname{\ Chiplets}^{\varname{Num. \ Layers}} \Big) \label{eqt:layer_placement} \\
\times \varname{\ Num. \ DNNs}! \label{eqt:schedule}
\end{gather}

Assuming fixed-dataflow \emph{SAT}s, term~\eqref{eqt:nmappings} is the mapping space size calculated by multiplying the number of loop permutations in DRAM (assuming these are not fixed) by the number of possible factorisation of each workload dimension. To calculate the latter, loop tiling can be thought of as the task of assigning a tiling level to each prime factor of the workload dimensions involved. Without spatial unrolling, the number of tiling levels coincides with the number of memory levels; otherwise, it will be greater. In term~\eqref{eqt:nmappings}, $7$ is the number of tensor dimensions in a convolutional layer, $m$ is the number of tiling levels (greater or equal to the number of levels in the \emph{SAT} memory hierarchy), and $F_d$ is the set of the prime factors of the $d^{th}$ workload dimension.

For a multi-accelerator system consisting of \emph{Chiplets} chiplets, term~\eqref{eqt:chiplet_template} represents the number of possible combinations in which \emph{SAT}s can be assigned to the chiplets.

Term~\eqref{eqt:param_space} represents the number of possible combinations of the various instances of \emph{SAT}s that can be assigned to the chiplets. As mentioned above, here it is assumed that \emph{Num. Params} and \emph{Num. Values} are uniform across all the \emph{SAT}s. However, in the generic case, each parameter \emph{p} will vary within a specific set with a cardinality of \emph{Num. Values(p)}. Hence, for the generic case, term~\eqref{eqt:param_space} will be modified as:
\begin{gather}
\left( \prod_{p=1}^{\varname{Num. \ Params}} \varname{Num.\ Values(p)} \right)^{\varname{Chiplets}}
\end{gather}
Term~\eqref{eqt:chiplet_placement} represents the possible combinations of assigning a chiplet to an NoP tile. Term~\eqref{eqt:layer_placement} represents the number of possible combinations for assigning a layer to a chiplet. Finally, term~\eqref{eqt:schedule} gives a lower bound on the possible schedules considering the combinations of execution order at the DNN granularity rather than the layer granularity.

The simulation configuration used to evaluate MOHaM in Section~\ref{sec:experiments} has \emph{Num. Layers} as 200 (average), \emph{Num. DNNs} as 4, \emph{Max. Chiplets} as 8, \emph{Num. SATs} as 3, \emph{Num. Params} as 4 (average), and \emph{Num. Values} as 4096 (average). Hence, the search space size is in the order of $10^{4318}$. To get an idea of how large this is, consider that the number of particles in the observable universe is estimated to be around $10^{80}$.




\begin{algorithm}
\footnotesize
\SetAlgoLined\DontPrintSemicolon
\SetKwFunction{proca}{LayerMapper}
\SetKwFunction{procb}{GlobalScheduler}
\SetKwFunction{procc}{MOHaM}
\SetKwFunction{procd}{UniqueLayers}
\SetKwFunction{proce}{RunMEDEA}
\SetKwFunction{procf}{Add}
\SetKwFunction{procg}{InitialPopulation}
\SetKwFunction{proch}{ApplyCrossoverOperators}
\SetKwFunction{proci}{ApplyMutationOperators}
\SetKwFunction{procj}{Evaluate}
\SetKwFunction{prock}{FastNonDominatedSort}
\SetKwFunction{procl}{Survival}
\SetKwFunction{procm}{GetParetoIndividuals}

\SetKwProg{myproc}{Procedure}{}{}

\myproc{\proca{AM, SSAT}}{
    MG$\gets\{\}$ \\
    \For{$layer$ in \procd{AM}}{
        ML$\gets\{\}$ \\
        \For{$arch$ in SSAT}{
            MF$\gets$\proce{$layer$, $arch$} \\
            ML.\procf{MF}
        }
        MG.\procf{ML}
    }
    \KwRet MG
}
\vspace{0.2cm}
\myproc{\procb{AM, SSAT, MG}}{
    PP$\gets$\procg{} \\
    \For{$g \gets 1$ to $G$}{
        \textcolor{blue}{\tcp{G = number of generations}}
        OP$\gets$\proch{PP} \\
        OP.\proci{OP} \\
        OP.\procj{} \\
        MP$\gets$\prock{PP, OP} \\
        OP$\gets$\procl{MP} \\
    }
    \KwRet PP.\procm{}
}
\vspace{0.2cm}
\myproc{\procc{AM, SSAT}}{
    MG$\gets$\proca{AM, SSAT} \\
    ST$\gets$\procb{AM, SSAT, MG} \\
    \KwRet ST \textcolor{blue}{\tcp{Pareto set of schedules}}
}
\caption{MOHaM high-level flow.}
\label{alg:moham}
\end{algorithm}
\section{The MOHaM Framework}

\begin{table}
\centering
\caption{Acronyms, abbreviations and key terms.}
\label{tab:terminology}
\resizebox{\columnwidth}{!}{
\begin{tabular}{@{}>{\bfseries}p{0.15\columnwidth}p{\dimexpr0.91\columnwidth-4\tabcolsep}@{}}
\toprule
AM & Application Model; a set of independent DNNs. \\
\addlinespace[0.1cm]
SAT & Sub-Accelerator Template; a parameterised and reconfigurable DNN accelerator with a fixed memory hierarchy. \\
\addlinespace[0.1cm]
SAI & Sub-Accelerator Instance of an SAT with fixed parameters. \\
\addlinespace[0.1cm]
MI & Memory Interface; connects memory banks with SAIs. \\
\addlinespace[0.1cm]
NoP & Network-on-Package; interconnect SAIs and MIs. \\
\addlinespace[0.1cm]
MAS & Multi-Accelerator System; a set of SAIs with memory. \\
\addlinespace[0.1cm]
S & Schedule; spatial and temporal allocation of each layer of every DNN of an AM on the SAIs of a MAS. \\
\addlinespace[0.1cm]
$\mathbf{M_{l,f,i}}$ & $i^{th}$ Pareto mapping for a layer $l$ in an SAT $f$. \\
\addlinespace[0.1cm]
$\mathbf{MF_{l,f}}$ & Pareto set of mappings for a layer $l$ in an SAT $f$. \\ 
\addlinespace[0.1cm]
$\mathbf{ML_{l}}$ & Pareto set of mappings for a layer $l$ in all the SATs. \\
\addlinespace[0.1cm]
MG & Pareto set of mappings for all the layers in all the SATs. \\
\addlinespace[0.1cm]
NSGA-II & Non-Dominated Sorting Genetic Algorithm~\cite{deb2002fast}. \\
\addlinespace[0.1cm]
Gene & A tuple of encoded values for a layer or an SAI. \\
\addlinespace[0.1cm]
Genome & An array of genes that represent all the information about a layer or an SAI, like mapping, execution order, etc. \\
\addlinespace[0.1cm]
Chromosome & A concatenation of genomes of all the layers and SAIs. \\
\addlinespace[0.1cm]
Individual & A chromosome with valid mapping, SAI and execution order of the layers, and SAT and NoP tiles for the SAIs. \\
\addlinespace[0.1cm]
Population & A set of individuals evolving through GA generations. \\
\addlinespace[0.1cm]
Pareto \break Efficiency & An individual is Pareto-efficient if no one else in the population is better for all the objectives at the same time. \\
\addlinespace[0.1cm]
Topological Sort & A linear ordering of a Directed Acyclic Graph (DAG) where a node appears before all the nodes it points to. \\
\bottomrule
\end{tabular}
}
\end{table}

To search such an enormous space, the proposed MOHaM framework adopts a two-step approach, \textit{layer mapping} and \textit{global scheduling}. Algorithm~\ref{alg:moham} presents a high-level flow of MOHaM, and the following sections describe its working. Table~\ref{tab:terminology} presents the acronyms, abbreviations and key terms.

\subsection{Layer Mapper}~\label{sec:lm}
In this step, each layer of the \emph{AM} is mapped onto each of the \emph{SAT}s from the input library. At all times, a layer is mapped onto any single instance of an \emph{SAT} and can not be split across multiple \emph{SAT} instances. Two layers, $l_i$ and $l_j$ can be instances of the same workload, i.e., have the same problem dimensions. Hence, to reduce the layer mapping search space, MOHaM only maps the unique layers to \emph{SAT}s.

Let $M_{l,f,i}$ be the $i^{th}$ Pareto mapping for a layer $l$ of the \emph{AM} in a \emph{SAT} $f$ from the input library. $M_{l,f,i}$ is a choice of a specific tiling, loop ordering, parallelism and clustering in $f$. Let $MF_{l,f}$ be the Pareto set of mappings with respect to the objectives for $l$ in $f$. $MF_{l,f}$ can be represented by: 
\begin{equation}
    MF_{l,f} = \{\,M_{l,f,i} \mid i = 0, \dots, (m_{l,f}-1)\,\}
    \label{eq:mf}
\end{equation}
where, $m_{l,f}$ is the total number of Pareto mappings. Now, let $ML_{l}$ be the Pareto set of mappings for $l$ in all the \emph{SAT}s of the input library. $ML_{l}$ can be represented by:
\begin{equation}
    ML_{l} = \{\,MF_{l,f} \mid f = 0, \dots, (F-1)\,\}
    \label{eq:ml}
\end{equation}
where, $F$ is the total number of \emph{SAT}s in the input library. Finally, let $MG$ be the Pareto set of mappings for all the layers of \emph{AM} in all the \emph{SAT}s. $MG$ can be represented by:  
\begin{equation}
    MG = \{\,ML_{l} \mid l = 0, \dots, (L-1)\,\}
    \label{eq:mg}
\end{equation}
where, L is the total number of layers in the \emph{AM}. MOHaM obtains the Pareto set of mappings in equation~\eqref{eq:mf} to~\eqref{eq:mg} by leveraging MEDEA~\cite{russo2022medea} (refer line 6 in Algorithm~\ref{alg:moham}). For each layer of the \emph{AM}, MEDEA searches for the Pareto set of mappings onto an \emph{SAT} instance. The Pareto set is for the chosen optimisation objectives, latency, energy and area. 

\subsection{Global Scheduler}
In this step, the Pareto set of mappings found for each layer in the previous step (Section~\ref{sec:lm}) is used to search the global scheduling. It returns a Pareto set of (\emph{MAS}, \emph{S}) pair with minimum latency, energy and area. This scheduler exploits one of the most widely accepted multi-objective GA, NGSA-II~\cite{deb2002fast}. It has five major phases, \textit{sampling}, \textit{selection}, \textit{crossover}, \textit{mutation} and \textit{survival}. While the original selection and survival phases are used, several custom genetic operators are designed for problem-specific crossover and mutation to increase sampling efficiency and decrease search time. 

\subsubsection{Chromosome Encoding}
\label{sec:chromosome_encoding}
\begin{figure}
    \centering
    \includegraphics[width=\columnwidth]{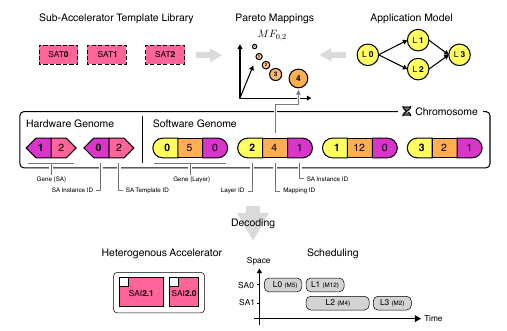}
    \caption{MOHaM global scheduler chromosome structure.}
    \label{fig:chromosome}
\end{figure}

\begin{figure*}
    \centering
    \includegraphics[width=\textwidth]{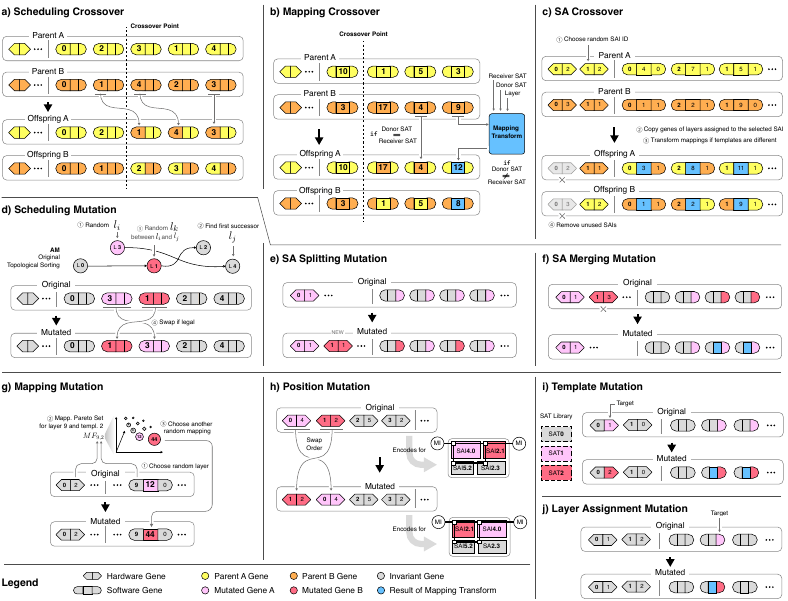}
    \caption{MOHaM-specific genetic operators.}
    \label{fig:operators}
\end{figure*}

A fundamental step in employing a GA is to define an encoding for the individuals in the population. MOHaM requires this encoding to represent, \textit{(a)} mapping strategy of each layer, \textit{(b)} \emph{SAI} of each layer, \textit{(c)} execution sequence of layers in the \emph{AM}, \textit{(d)} SA template of each \emph{SAI}, and \textit{(e)} NoP tile of each \emph{SAI}. A two-part chromosome used by the scheduler is shown in Figure~\ref{fig:chromosome} and is described here:
\begin{itemize}    
    \item \textbf{Hardware Genome:} It encodes the instances of the \emph{SAT}s. It is an array of genes, where each gene denotes an SAI. Each gene is a tuple $\langle SAI, SAT \rangle$ where \emph{SAI} is the SA instance and \emph{SAT} is the template identifier. The order of the genes represents the position of the \emph{SAI} hosting tile in the NoP. The number of genes is equal to the number of \emph{SAI}s and varies between 1 and maximum NoP tiles across chromosomes.
    \item \textbf{Software Genome:} It encodes the layers of the \emph{AM}. It is an array of genes, where each gene denotes a layer. Each gene is a tuple $\langle LID, MID, SAI \rangle$ where \emph{LID} is the layer identifier, \emph{MID} is the mapping identifier, and \emph{SAI} is the SA instance the layer will be executed. The order of the genes is a topological sorting of the layers using Kahn's algorithm~\cite{kahn1962topological} and represents the temporal sequence in which the layers will be executed. The number of genes is equal to the number of layers in the \emph{AM} and is fixed for chromosomes.
\end{itemize}

\subsubsection{Custom Genetic Operators}
The standard crossover swaps parental genes at a pivot point to generate offspring, while standard mutation randomly changes some genes of the offspring to introduce variation. However, directly applying these operators to MOHaM will not work since offspring chromosomes must adhere to specific constraints to be valid individuals. Each MOHaM chromosome is a tuple $\langle \mathit{genome}^{hw},\mathit{genome}^{sw} \rangle$ where $\mathit{genome}^{hw}$ can be represented by:
\begin{multline*}
    \mathit{genome}^{hw} = \{\,g_i^{hw} = \langle SAI, SAT \rangle \mid \\ i = 1, \dots, Chiplets\,\}
\end{multline*}
Whereas $\mathit{genome}^{sw}$ can be represented by:
\begin{multline*}
    \mathit{genome}^{sw} = \{\,g_j^{sw} = \langle LID, MID, SAI \rangle \mid \\ j = 1, \dots, Layers\,\}
\end{multline*}
The constraints for valid individuals are specified as follows:
\begin{itemize}
    \item[C1] In the hardware genome, the \emph{SAT} of a gene must refer to one of the templates from the input library.
    \begin{multline*}
      g_i^{hw} = \langle \bullet, SAT_i \rangle \mid SAT_i \in SAT \ Input \ Library
    \end{multline*}
    
    \item[C2] In the hardware genome, each \emph{SAI} must be referred to by at least one gene from the software genome.
    \begin{multline*}
        g_i^{hw} = \langle SAI_i, \bullet \rangle \mid \exists j \ with \ g_j^{sw} = \langle \bullet, \bullet, SAI_i \rangle
    \end{multline*}

    \item[C3] In the software genome, the \textit{MID} of a layer \textit{l} must refer to one of the mappings from $MF_{l,f}$, where \textit{f} is the template of the \emph{SAI} on which \textit{l} will be executed.
    \begin{multline*}
        g_j^{sw} = \langle \bullet, MID_j, SAI_j \rangle \mid MID_j \in MF_{l_j,f_j} \\
        where \ f_j = SA \ Template (SAI_j)
    \end{multline*}
    
    \item[C4] In the software genome, the \emph{SAI} of a layer must refer to one of the encoded genes in the hardware genome.
    \begin{multline*}
        g_j^{sw} = \langle \bullet, \bullet, SAI_j \rangle \mid \exists i \ with \ g_i^{hw} = \langle SAI_j, \bullet \rangle
    \end{multline*}
    \item[C5] In the software genome, the order of the genes must be a valid topological sorting that respects the dependencies between layers of the DNNs from the AM.
    \begin{multline*}
        \forall j,k = 1, \ldots, Layers \mid j < k \\
        g_j^{sw}=\langle LID_j, \bullet, \bullet \rangle, \ g_k^{sw}=\langle LID_k, \bullet, \bullet \rangle \\
        \Rightarrow (LID_k, LID_j) \notin D
    \end{multline*}
    where $D$ is the set of dependencies between layers.
\end{itemize}

Therefore, the choice of crossover and mutation operators significantly impacts the performance of a GA, and it is often necessary to adapt them to the specific problem at hand. While there is no systematic way, using domain-specific knowledge of the problem is a common practice for customising crossover and mutation. Hence, a set of custom crossover and mutation operators are designed by considering the problem-specific constraints to (a) explore the search space strategically, and (b) increase the sampling efficiency. They are shown in Figure~\ref{fig:operators} and described here:

\emph{(a) Scheduling Crossover}:
It combines the topological sorting of the parent chromosomes, as shown in Figure~\ref{fig:operators}(a). It generates offspring by taking the first part of one of the parents, i.e., all the genes before the crossover point, and appending all the unique genes from the other parent. Existing literature~\cite{xu2014genetic} has proof that by proceeding this way, the offspring will also have valid topological sorting.

\emph{(b) Mapping Crossover}:
It combines the mappings of the parent chromosomes, as shown in Figure~\ref{fig:operators}(b). It generates offspring by taking layer mappings from the first part of one of the parents, i.e., all the genes before the crossover point, and the remaining mappings from the other parent. However, a new mapping might be invalid if it is for an SA template different from that of the \emph{SAI} on which the layer will be executed. In such a case, a compensation mechanism called \emph{Mapping Transform} finds the most similar one among all the possible mappings for the layer in that \emph{SAI}.

\emph{(c) SA Crossover}:
It swaps a random \emph{SAI} at $i^{th}$ position, $sai_i$, between the parent chromosomes, as shown in Figure~\ref{fig:operators}(c). If both parents \emph{A} and \emph{B} have an \emph{SAI} at the $i^{th}$ position, they are swapped and two offspring are generated. In such a case, if $sai_i$ of \emph{A} and $sai_i$ of \emph{B} are of different SA templates, all the mappings from both of their layers undergo \emph{Mapping Transform}. Whereas, if only one parent has an \emph{SAI} at the $i^{th}$ position, it is added to the other parent with all its assigned layers and only one offspring is generated.

\emph{(d) Scheduling Mutation}:
It mutates the topological sorting of a chromosome, as shown in Figure~\ref{fig:operators}(d). Let $l_i$ be a random gene (layer) from the software genome. Let $l_j$ be the nearest layer in the traversal that is dependent on $l_i$. Let $l_k$ be a random layer between $l_i$ and $l_j$. If all the layers on which $l_k$ has a dependency lie before $l_i$ in the traversal, their position can be swapped. Similar to~\ref{fig:operators}(a), the mutated chromosome will also have a valid topological sorting~\cite{xu2014genetic}.

\emph{(e) SA Splitting Mutation}:
It reduces the load in a random \emph{SAI} at $i^{th}$ position, $sai_i$, to mutate a chromosome, as shown in Figure~\ref{fig:operators}(e). A new $sai_j$ of the same SA template is appended to the hardware genome. Thereafter, half of the layers currently assigned to $sai_i$ are randomly chosen and assigned to $sai_j$. The goal is to increase parallelisation.

\emph{(f) SA Merging Mutation}:
It increases the load in a random \emph{SAI} at $i^{th}$ position, $sai_i$, to mutate a chromosome, as shown in Figure~\ref{fig:operators}(f). Another existing \emph{SAI} at $j^{th}$ position, $sai_j$, is randomly chosen and all its layers are assigned to $sai_i$. If $sai_j$ is of a different SA template than $sai_i$, all the mappings of the imported layers undergo \emph{Mapping Transform}. The goal is to reduce chip area with reduced \emph{SAI}s.

\emph{(g) Mapping Mutation}:
It modifies the \emph{MID} of a random layer $l$ to mutate a chromosome, as shown in Figure~\ref{fig:operators}(g). All the possible mappings of $l$ are in the $MF_{l,f}$ Pareto set, where \textit{f} is the template of the \emph{SAI} on which \textit{l} will be executed. One of those possible mappings is assigned to $l$ for mutation.

\emph{(h) SA Position Mutation}:
It swaps the position of two \emph{SAI} hosting tiles in the 2D Mesh NoP, as shown in Figure~\ref{fig:operators}(h). The goal is to find a configuration where the system bandwidth is distributed among the \emph{SAI}s and \emph{MI}s in a way that avoids communication bottlenecks and reduces latency.

\emph{(i) SA Template Mutation}:
It modifies the SA template of a random \emph{SAI} to mutate a chromosome, as shown in Figure~\ref{fig:operators}(i). All the mappings from different layers to be executed on the mutated \emph{SAI} undergo \emph{Mapping Transform}.

\emph{(j) Layer Assignment Mutation}:
It modifies the \emph{SAI} of a random layer $l$ to mutate a chromosome, as shown in Figure~\ref{fig:operators}(j). If the modified \emph{SAI} is of a different SA template, the corresponding mapping for $l$ undergoes \emph{Mapping Transform}.

These operators are input agnostic, i.e., MOHaM can be used for any DNN as an \emph{AM}, with any accelerator as an \emph{SAT}, without modifying or adding any custom GA operator.

\subsection{Objectives Evaluation}
Three objective metrics are evaluated for each individual: (a) \textit{latency}, which is the makespan of the workload, (b) \textit{energy}, which is the total energy consumption to execute the workload, and (c) \textit{area}, which is the real estate of the hardware for the workload. They depend on the system configuration and the schedule. For example, increasing the number of \emph{SAI}s decreases latency but increases area. Similarly, the mapping strategy for each layer of the \emph{AM}, and the \emph{SAT}s affect all three objectives. MOHaM translates the chromosome encoding into objective metrics using the Timeloop~\cite{parashar2019timeloop} + Accelergy~\cite{wu2019accelergy} cost model (refer Figure~\ref{fig:moham}).

The translation begins from the hardware genome. Each of its genes is converted into an \emph{SAI} of the appropriate \emph{SAT}. At this point, parameter values of the \emph{SAI}s are unknown. Then software genomes are examined to identify which mappings are used for each gene (layer). All the free parameters, including the number of PEs, size of buffers, etc., are set to the maximum required by the mappings for all the layers assigned to an \emph{SAI}. For example, Figure~\ref{fig:moham}(e) shows that $L2.0$ and $L1.1$ are assigned to \emph{SAI0.0}. If they require 64 and 128 PEs and 64 KiB and 32 KiB shared buffers, respectively, \emph{SAI0.0} will be assigned 128 PEs and 64 KiB shared buffer. Each \emph{SAI} executes its assigned layers, and the total area is estimated. This over-provision is automatically handled, as MOHaM finally identifies individuals with minimum area.

As each mapping has its corresponding energy consumption, summing them for all the layers could provide the total energy consumption. However, energy for reads and writes in the buffers depends on their sizes. Hence, energy for each mapping must be estimated based on the updated values of the free parameters in the \emph{SAI}s.

Latency is estimated using the topological sorting encoded in the software genome. Each of its genes (layers) is read sequentially and mapped on the assigned \emph{SAI}s. This traversal guarantees that a layer's dependencies are already scheduled before its turn comes. Each layer has a start and end time, and the total latency is estimated based on the latest end time among all the layers. However, this is true only when no communication or memory bottlenecks exist.

\subsubsection{Communication Modeling} \label{sec:communication}
As shown in Figure~\ref{fig:moham}(d), MOHaM assumes that a 2D Mesh NoP interconnects the \emph{SAI}s and memory banks. The \emph{SAI}s communicate with the memory banks through the \emph{MI}s. The NoP is implemented via on-package links in a passive silicon interposer. It employs efficient intra-package signalling circuits using Ground-Reference Signaling (GRS) technology. Specifically, each chiplet has eight chiplet-to-chiplet GRS transceivers; four transmitters and four receivers. A transceiver has four data lanes, each providing 4 GB/s, thus a total peak chiplet bandwidth of 4 * 4 GB/s = 16 GB/s and energy of 0.82 pJ/bit~\cite{shao2019simba}. The NoP employs the XY routing algorithm. There is no inter-chiplet communication and the \emph{SAI}s only communicate with the \emph{MI}s. While the positions of the \emph{MI}s are configurable, MOHaM assumed one at each corner of the 2D Mesh NoP for the evaluation in Section~\ref{sec:experiments}. \emph{SAI}s reads and writes in their nearest \emph{MI}s. Some CPU processing happens between layer execution (e.g. tensor reordering), and the output is stored in the nearest memory bank of the \emph{SAI} executing the next layer. Depending on their position in the NoP, multiple \emph{SAI}s could be assigned the same \emph{MI}, thus competing for the shared link. Let 16 \emph{SAI}s be arranged in a 4$\times$4 2D Mesh NoP with 4 \emph{MI}s, one at each corner. Here, 4 \emph{SAI}s will be sharing 1 \emph{MI} and the NoP link connecting that \emph{MI} will be the bottleneck link. So, either the shared \emph{MI} bandwidth ($BW_{MI}$) or the bottleneck link bandwidth ($BW_{BL}$) will be the reason for the congestion.

Each layer has a certain read and write bandwidth requirement when executed on its assigned \emph{SAI}. If a time segment has parallel execution of layers and the combined bandwidth requirement of the layers executed on the \emph{SAI}s sharing the same \emph{MI} exceeds either the $BW_{MI}$ or the $BW_{BL}$, there will be stalls. Let the time segment be $n$ cycles and the combined bandwidth requirement be $BW_{req}$. The stalled time segment of $m$ cycles can be calculated by:
\begin{equation*}
    m = n \times \frac{BW_{req}}{min(BW_{MI},BW_{BL})}
    \label{eq:comm_latency}
\end{equation*}

\noindent $m>n$ signifies a delay in the execution of certain layers. In such cases, MOHaM compensates the start times of all the subsequent layers with dependencies on the delayed layers. However, if a subsequent layer is executed on an \emph{SAI} that communicates with a different \emph{MI}, the delay in its start time might create a new stalled time segment. MOHaM considers all such cases to correctly estimate the latency. After evaluating all the stalled time segments, the latest end time among all the layers becomes the new total latency.

To estimate the communication energy, MOHaM considers the amount of data transferred between an \emph{SAI} and its nearest \emph{MI} and the hop distance between them following the XY routing algorithm. It can be calculated by:
\begin{multline*}
    \sum_{\varname{i}=1}^{\varname{Num. \ SAIs}} \Big( \varname{Bits. \ Transferred}_{\varname{SAI_i}} \\
    \times \ \varname{Hops}_{\varname{XY}}(\varname{SAI_i},\varname{MI}) \\
    \times \ \varname{Energy/Hop/Bit} \Big)
    \label{eq:comm_energy}
\end{multline*}


The NoP real estate is not currently considered in the area estimation. MOHaM considers a silicon interposer-based NoP which is outside the chiplet areas. Hence, it is ignored while estimating the area. Nevertheless, it can be easily included in the estimation by multiplying the number of \emph{SAI}s with a NoP router area. Depending on the number of ports, different router areas can also be considered.

\subsubsection{Convergence Criterion}
The proposed MOHaM framework supports a GA stopping criterion based on the density of the non-dominated solutions presented in~\cite{roudenko2004steady}. Alternatively, simulating for a fixed number of generations can also be configured very easily.
\begin{table}
\centering
\caption{Multi-DNN workload scenarios.}
\label{tab:workloads}
\resizebox{0.8\columnwidth}{!}{
\begin{tabular}{@{}cccc@{}}
\toprule
\textbf{\#} & \textbf{Workload} & \textbf{Domain} & \textbf{DNN Model} \\
\midrule
\multirow{4}{*}{A} & \multirow{4}{*}{AR/VR} & Image Classification & ResNet50 \\
&& Image Segmentation & UNet \\
&& Object Detection & SSD-MobileNetV1 \\
&&& YOLOv3 \\
\midrule
\multirow{3}{*}{B} & \multirow{3}{*}{Edge} & Image Classification & ResNet50 \\
&& Object Detection & SSD-ResNet34 \\ 
&& Language Processing & BERT-Large \\
\midrule
\multirow{3}{*}{C} & \multirow{3}{*}{Mobile} & Image Classification & MobileNetV3L \\
&& Image Segmentation & DeepLabV3+ MN2 \\
&& Language Processing & Mobile-BERT \\
\midrule
\multirow{4}{*}{D} & \multirow{4}{*}{Data Center} & Image Classification & GoogleNet \\
&& Object Detection & YOLOv3 \\
&& Language Processing & BERT-Large \\
&& Recommendation & DLRM \\
\bottomrule
\end{tabular}
}
\end{table}

\section{Evaluation}\label{sec:experiments}
This section presents multiple experiments conducted to evaluate the performance of the MOHaM framework.

\subsection{Methodology}
\subsubsection{DNNs and Workloads}
MOHaM is to be used at design-time of a multi-accelerator system, where the application domains will be known or guessed apriori. Hence, AR/VR, edge, and mobile workloads are considered for evaluation. Besides, standard data center workloads are dominated by vision, language and recommendation-based DNNs~\cite{anderson2021first}. Hence, MOHaM is also evaluated with such a workload to show its effectiveness if the application domains can be guessed. All of them are multi-DNN workloads created with models from the MLPerf benchmark suite~\cite{reddi2020mlperf}\cite{mattson2020mlperf}. Table~\ref{tab:workloads} presents the workload scenarios along with their DNN models. MOHaM takes them as inputs in the ONNX~\cite{onnx} interoperable format.

\subsubsection{Sub-Accelerator Templates}
The following state-of-the-art constitutes the \emph{SAT} library\footnote{Any accelerator can be added in the template library of MOHaM.}:

\begin{itemize}
    \item Eyeriss~\cite{chen2016eyeriss}: Row-stationary dataflow
    \item Simba~\cite{shao2019simba}: Weight-stationary dataflow
    \item ShiDianNao~\cite{du2015shidiannao}: Output-stationary dataflow
\end{itemize}

They are heterogeneous accelerators and are chosen to support diverse DNNs. Table~\ref{tab:mohamconfig} presents MOHaM configuration for the experiments, where the GA exploration parameters are based on the guidelines of the widely adopted~\cite{eiben2011parameter}, and the architectural parameters are based on the state-of-the-art~\cite{chen2016eyeriss}\cite{shao2019simba}\cite{du2015shidiannao}. For the parameter names beginning with ``\emph{Max.}'', the presented values are their upper bound and are reconfigurable. For a chosen Pareto solution, two \emph{SAI}s of the same \emph{SAT}, say Eyeriss, can have different values for the number of PEs, shared buffer size and PE scratchpad size. The upper bound is to limit the search space for exploration.

\begin{table}[t]
\centering
\caption{MOHaM simulation configuration.}
\label{tab:mohamconfig}
\resizebox{\columnwidth}{!}{
\begin{tabular}{@{}rlrl@{}}
\toprule
\multicolumn{4}{c}{\textbf{Exploration Parameters}} \\
\midrule
Num. Generations & 300 &
Population Size & 250 \\
Sched. Cross. Prob. & 0.103 &
Sched. Mut. Prob. & 0.052 \\
SA Cross Prob. & 0.045 &
Template Mut. Prob. & 0.041 \\
Merging Mut. Prob. & 0.042 &
Splitting Mut. Prob. & 0.039 \\
Mapping Mut. Prob. & 0.048 &
Mapping Cross. Prob. & 0.047 \\
Layer Assign. Mut. Prob. & 0.025 &
Position Mut. Prob. & 0.027 \\
Max. SA Instances & 8  & \\ 
\midrule
\multicolumn{4}{c}{\textbf{Common Architecture Parameters}} \\
\midrule
Technology Node & 45 nm &
DRAM Technology & LPDDR4 \\
Mem. Interface BW & 4 GB/s &
Clock Frequency & 1 GHz \\
Word Size & 8 bits &
SRAM Buf. BW & 16 GB/s \\
\midrule
\multicolumn{4}{c}{\textbf{Eyeriss-like Template}} \\
\midrule
Dataflow & Row-Stat. & 
Max. Num. of PEs & 4096 \\
Max. Shared Buf. Size & 131 KiB &
Max. PE Scratchpad Size & 0.5 KiB \\
\midrule
\multicolumn{4}{c}{\textbf{Simba-like Template}} \\
\midrule
Dataflow & Weight-Stat. &
Max. Num. of PEs & 64 \\
Max. MACs per PE & 64 &
Max. Global Buf. Size & 64 KiB  \\
Max. Weight Buf. Size & 32 KiB &
Max. Input Buf. Size & 8 KiB \\
Max. Accum. Buf. Size & 3 KiB & \\
\midrule
\multicolumn{4}{c}{\textbf{ShiDianNao-like Template}} \\
\midrule
Dataflow & Output-Stat. &
Max. Num. of PEs & 4096 \\
Max. Neurons Buf. Size & 131 KiB &
Max. Synapses Buf. Size & 131 KiB \\
\bottomrule
\end{tabular}
}
\end{table}

\subsubsection{Solution Anatomy}

\begin{figure*}
    \centering
    \includegraphics[width=\textwidth]{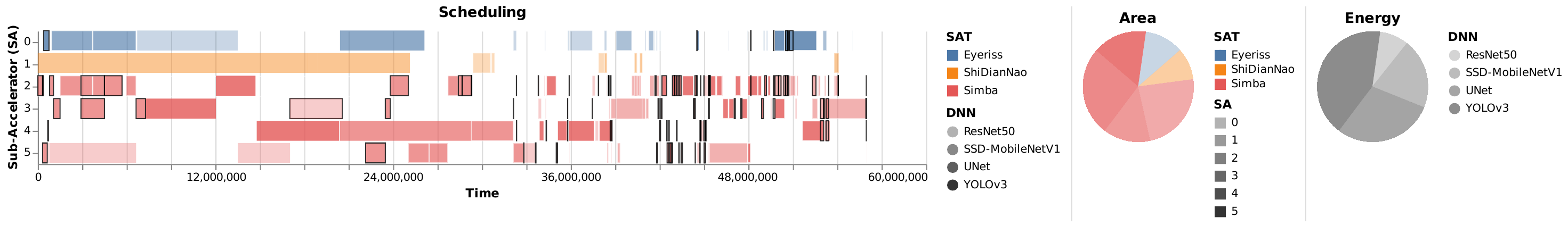}
    \caption{Scheduling Gantt chart for latency, pie charts for area and energy breakdowns of a Pareto solution.}
    \label{fig:solution}
\end{figure*}

\begin{figure*}
    \centering
    \includegraphics[width=\textwidth]{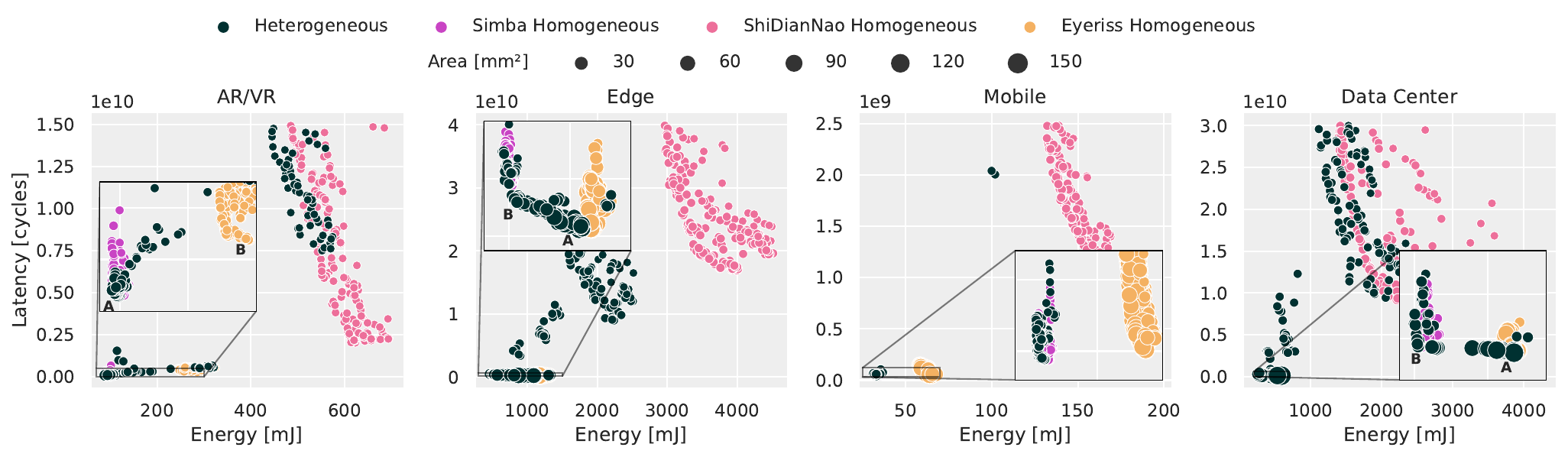}
    \caption{Comparison of Pareto solutions with homogeneous and heterogeneous sub-accelerators.}
    \label{gra:hevsho}
\end{figure*}

Figure~\ref{fig:solution} shows the visual solution anatomy available in MOHaM. It shows the scheduling Gantt chart for latency, and two pie charts for energy and area of a Pareto solution from the AR/VR workload. The Gantt chart shows, on the y-axis, the \emph{SAI}s, and on the x-axis, the start and end times of the execution of each layer, measured in cycles. Each bar represents the execution of a layer of the \emph{AM} on a particular \emph{SAI}. Black traces represent bandwidth-constrained execution segments as described in Section~\ref{sec:communication}. The latest end time among all the layers is the latency (makespan) of the workload. The area pie chart displays each \emph{SAI}'s contribution to the real estate of the \emph{MAS}. Instances from the same \emph{SAT} have the same colour. The energy pie chart displays each DNN's contribution to the total energy consumption to execute the workload. The Pareto set might have another solution with better scheduling and silicon utilisation. The solution anatomy helps in comparing design points.

\subsection{Results}
\begin{figure*}
    \centering
    \includegraphics[width=\textwidth]{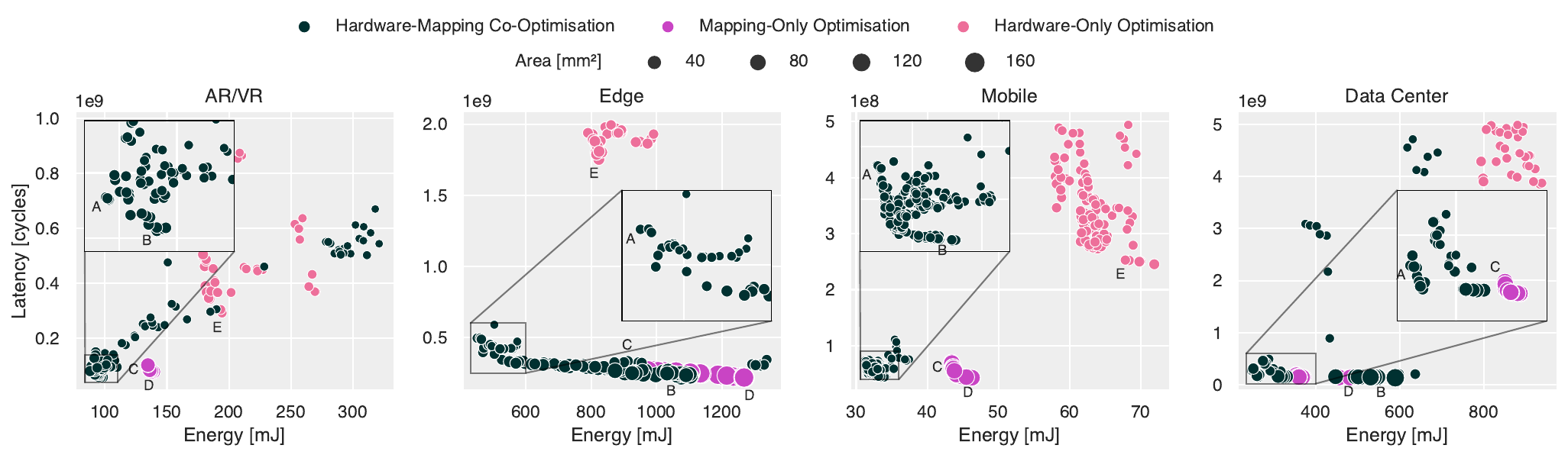}
    \caption{Comparison of Pareto solutions with hardware-only, mapping-only, and hardware-mapping co-optimisation.}
    \label{gra:opvsco}
\end{figure*}

\begin{figure*}
    \centering
    \includegraphics[width=\textwidth]{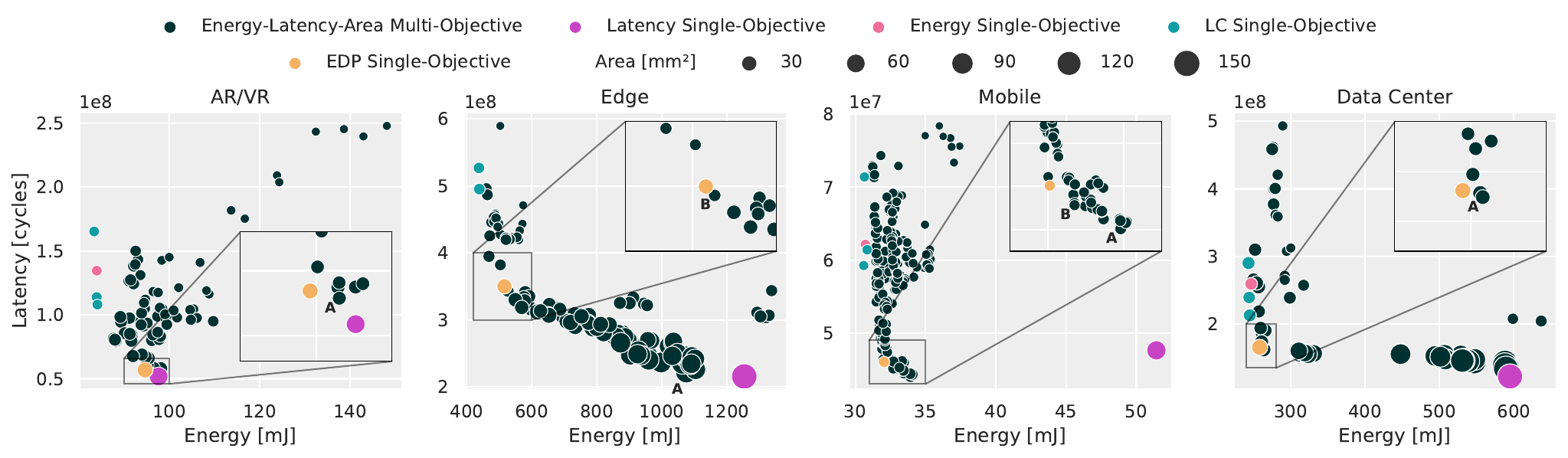}
    \caption{Comparison of individual solutions with mono-objective and Pareto solutions with multi-objective exploration.}
    \label{gra:sivsmu}
\end{figure*}

For all the results shown in Figures~\ref{gra:opvsco},~\ref{gra:hevsho},~\ref{gra:sivsmu} and~\ref{gra:mohamvssota}, the x-axis represents energy in \textit{mJ}, the y-axis represents latency in \textit{cycles}, and the size of the design points represents the area in \textit{mm\textsuperscript{2}}. A common observation valid for all these results is the distribution of the solutions found by MOHaM. They are spread over a large Pareto surface rather than limited to a specific region. This is an important advantage of MOHaM as it provides a variety of trade-off solutions to choose from.

\subsubsection{Homogeneous vs Heterogeneous Accelerators}
This experiment evaluates the need for heterogeneous sub-accelerators (SAs) in multi-accelerator systems. Figure~\ref{gra:hevsho} shows the comparison of Pareto solutions with homogeneous and heterogeneous SAs. For homogeneous SAs, MOHaM is run once, each with only Eyeriss-like, Simba-like, and ShiDianNao-like SA templates. Then, they are compared with the result of a complete MOHaM run for heterogeneous SAs. It is observed that for the \textit{Edge} and \textit{Data Center} workloads, Eyeriss (yellow) has lower latency at the cost of higher energy and bigger area. On the contrary, ShiDianNao (red) is area efficient but at the cost of very high energy and latency. For the \textit{AR/VR} and \textit{Mobile} workloads, Simba (purple) has the best energy and latency, while ShiDianNao (red) has the worst latency, respectively. Among the solutions with homogeneous SAs, Simba (purple) generally has better energy, while ShiDianNao (red) has a better area. With heterogeneous SAs, the solutions (black) are more uniformly distributed. For the \textit{AR/VR} workload, they have solutions with the lowest latency, energy, and area. For example, design point \emph{A} from heterogeneous (black) can achieve a $89.4\%$ latency reduction compared to design point \emph{B} from Eyeriss (yellow). Heterogeneous SA-based solutions are equally good for other workloads. For example, for the \textit{Data Center} and \textit{Edge} workloads, design point \emph{A} from heterogeneous (black) can achieve a $9.72\%$ and $20.41\%$ latency reduction, respectively, compared to design point \emph{B} from Simba (purple). Besides, the lowest energy solutions by heterogeneous SAs can achieve a $24.7\%$ reduction compared to Simba. This is possible as layers with specific dataflow preferences can be executed on the appropriate SAs. MOHaM instantiates the right set of SAs from the available templates and maps the layers for efficient execution. \emph{The key takeaway from this experiment is that flexible dataflow support with heterogeneous SAs increases the scalability of a multi-accelerator system toward diverse and emerging workloads}.

\subsubsection{Independent vs Simultaneous Optimisation}
This experiment evaluates the need for hardware-mapping co-optimisation in multi-accelerator systems. Figure~\ref{gra:opvsco} shows the comparison of Pareto solutions with hardware-only, mapping-only and hardware-mapping co-optimisation. For hardware-only optimisation, MOHaM is run with only Simba-like SA templates to have a fixed dataflow (e.g., weight-stationary), similar to ConfuciuX~\cite{kao2020confuciux}. For mapping-only optimisation, MOHaM is run with a fixed hardware configuration of eight heterogeneous SAs, similar to MAGMA~\cite{kao2022magma}. Finally, they are compared with the result of a complete MOHaM run for hardware-mapping co-optimisation. It is observed that for the \textit{AR/VR} workload, hardware-only optimisation (red) has a smaller area but high latency and energy. This is due to the fixed mapping (dataflow) strategy for all the layers. Whereas mapping-only optimisation (purple) has lower latency and energy but costs a bigger area. This is due to the fixed hardware configuration. In general, hardware-only optimisation has a smaller area but poor latency and energy across all workloads. Similarly, mapping-only optimisation has lower latency but a big area. With hardware-mapping co-optimisation, the solutions (black) are more uniformly distributed and achieve better performance. For example, for the \textit{AR/VR} workload, design point $A$ from co-optimisation (black) has two times the area as design point $E$ from hardware-only optimisation (red) but has $55\%$ less energy and $72\%$ less latency. It is also observed that mapping-only optimisation achieves similar latency but with higher energy and a bigger area. For example, for the \textit{Edge} workload, design point $B$ from co-optimisation (black) has the same latency as design point $D$ from mapping-only optimisation (purple) but has $15.3\%$ less energy and $36.5\%$ less area. Similar solutions exist in \textit{Mobile} and \textit{Data Center} workloads. This results from MOHaM instantiating the SAs with optimal hardware resources and mapping the layers per their dataflow preferences for optimal execution. \emph{The key takeaway from this experiment is that hardware-mapping co-optimisation can support diverse workloads with the best overall performance in a multi-accelerator system}.

\subsubsection{Single-Objective vs Multi-Objective Exploration}
This experiment evaluates the need for multi-objective exploration in multi-accelerator systems. Figure~\ref{gra:sivsmu} shows the comparison of individual solutions with mono-objective and Pareto solutions with multi-objective exploration. 
For mono-objective exploration, MOHaM is run once, each with only latency, energy, and EDP as an objective. Three Linear Combinations (LCs) of energy and latency with coefficients: $(0.25,0.75)$, $(0.5,0.5)$, and $(0.75,0.25)$ are also run once.
Then, they are compared with the result of a complete MOHaM run for multi-objective exploration. When objectives conflict, the best solution for one may sometimes lead to the worst for others with mono-objective exploration. For example, for the \textit{Data Center} workload, the solution with energy as an objective (red) is at the extreme left (i.e., best), but the latency is relatively high. Similarly, the solution with latency as an objective (purple) is best, while the area is worst. Even a slightly better solution for one objective may sometimes lead to heavily penalising others with mono-objective exploration. For example, for the \textit{Edge} workload, the design point with latency as an objective (purple) is only $3\%$ better than design point \emph{A} from multi-objective (black), but comes at the cost of $14.2\%$ more energy and $30.1\%$ more area. Furthermore, the best solution for one objective may sometimes be sub-optimal with mono-objective exploration. For example, for the \textit{Mobile} workload, the design point with latency as an objective (purple) has $7.5\%$ more latency, $34\%$ more energy and $75\%$ more area than design point \emph{A} from multi-objective (black). Even the best solution for popular aggregated objectives, EDP, may sometimes be sub-optimal with mono-objective exploration. For example, compared to the design point with EDP as an objective (yellow), multi-objective (black) design points \emph{A}, \emph{B}, \emph{B} and \emph{A} from the \textit{AR/VR}, \textit{Edge}, \textit{Mobile} and \textit{Data Center} workloads have $31.78\%$, $45.09\%$, $5.21\%$ and $20.49\%$ less area, respectively, while achieving the same EDP ($\pm0.04\%$). Similarly, the other aggregated objectives, LCs, provide unbalanced solutions towards energy despite having different coefficients. This behaviour demonstrates that choosing coefficients in a weighted sum objective can be challenging and time-consuming when searching for compromise solutions. MOHaM supports distinct multi-objective exploration and provides a Pareto set of solutions (black). Across all workloads, it provides multiple competitive solutions with the lowest energy, latency and area, and exposes interesting trade-offs that can be leveraged. \emph{The key takeaway from this experiment is that multi-objective exploration helps to identify the most suitable design for a specific multi-accelerator system}.

\begin{figure}
    \centering
    \includegraphics[width=0.7\columnwidth]{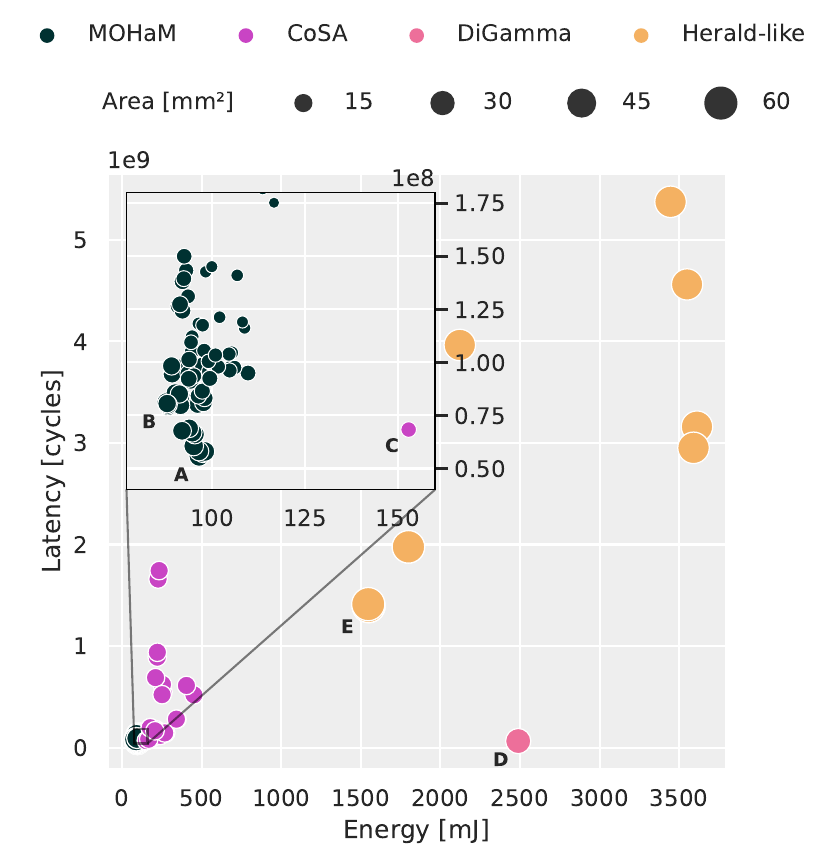}
    \caption{Comparison with state-of-the-art.}
    \label{gra:mohamvssota}
\end{figure}

\subsubsection{Comparison with State-of-the-Art}
This experiment evaluates MOHaM-generated multi-accelerator system designs against three popular state-of-the-art, CoSA~\cite{huang2021cosa}, DiGamma~\cite{kao2022digamma} and Herald~\cite{kwon2021heterogeneous}. The open-source implementations of CoSA and DiGamma are considered while Herald is implemented by faithfully following its evaluation settings. CoSA is a mono-objective, mapping-only optimisation that exploits Mixed-Integer Programming (MIP) to generate optimal mapping strategy without requiring iterative scheduling leveraging the Timeloop~\cite{parashar2019timeloop} cost model. DiGamma is a mono-objective, hardware-mapping co-optimisation that exploits GA (similar to MOHaM) for faster DSE leveraging the MAESTRO~\cite{kwon2020maestro} cost model. Herald is a mono-objective, hardware-mapping co-optimisation that employs an exhaustive search and heuristic-based scheduling for DSE of multi-accelerator systems leveraging the MAESTRO cost model.
Among the existing works on multi-accelerator systems, Herald is the only one targeting heterogeneous SAs, multi-DNN workloads and hardware-mapping co-optimisation (refer Table~\ref{tab:sota}). However, there are a few important differences between Herald and MOHaM:
\begin{itemize}
    \item Herald optimises the partitioning of the MAC budget among a fixed number of heterogeneous SAs. Whereas, MOHaM not only allows exploring the number of optimal SAs but also their buffer sizes.
    \item Herald uses a manual-tuned mapper that lacks flexibility and limits its scalability to diverse accelerator templates. Whereas, MOHaM exploits GA for its mapper that does not require tuning or modification with any accelerator template or workload variation.
    \item Herald provides Pareto optimal solutions for a mono-objective DSE. Whereas, MOHaM offers Pareto optimal solutions for a multi-objective DSE.
\end{itemize}
Herald originally employed the MAESTRO cost model, but to have a fair comparison with MOHaM, the Timeloop cost model is used here. Hence the resultant implementation is named \emph{Herald-like} (and not exactly Herald). Additionally, the number of MACs and SAIs equal to that in the maximum configuration of MOHaM is considered for Herald-like.
CoSA and DiGamma simulations are conducted considering flexible-dataflow architectures without considering the reconfiguration overheads. For CoSA, a Simba-like architecture is considered with the number of MACs equal to that in the maximum configuration of MOHaM. Its objective function is a linear combination of three components, \emph{compute}, \emph{traffic}, and \emph{buffer utilisation}. Hence, it is run with different coefficients of these components for multiple solutions. For DiGamma, even though it originally employed the MAESTRO, an equivalent implementation with the Timeloop is also available and is used here. As DiGamma operates in a layer-wise manner, the results of all the layers from the workload are combined, and the area budget is set to the maximum configuration of MOHaM.

\begin{table}
\centering
\caption{Configuration of design points from Figure~\ref{gra:mohamvssota}.}
\label{tab:sota_details}
\resizebox{\columnwidth}{!}{
\begin{tabular}{@{}lccccc@{}}
\toprule
& \multicolumn{5}{c}{\textbf{Design Points}} \\
\cmidrule(lr){2-6} 
\textbf{Metrics} & \textbf{A} & \textbf{B} & \textbf{C} & \textbf{D} & \textbf{E} \\
\midrule
Latency [$10^6$ cycles] & 55.76 & 80.59 & 68.33 & 64.12 & 1391.44 \\
\ \ Memory Bound & 64.69\% & 17.28\% & 100\% & N.A. & 35.43\% \\
\midrule
Energy [mJ] & 96.53 & 87.93 & 152.91 & 2488.94 & 1547.88 \\
\midrule
Area [mm$^2$] & 39.95 & 32.86 & 20.69 & 39.33 &  66.34 \\
\ \ Compute Area & 37.06\% & 36.25\% & 52.61\% & 17.25\% & \phantom{0}5.68\% \\
\ \ PE Buf. Area & 45.04\% & 43.12\% & 45.37\% & 71.85\% & 23.34\% \\
\ \ Shared Buf. Area & 17.90\% & 20.63\% & \phantom{0}2.02\% & 10.90\% & 70.98\% \\
\midrule
Total MACs & 14336 & 16384 & 32768 & 32768 & 9984\\
\midrule
Total SA Instances & 7 & 8 & 1 & 1 & 3 \\
\ \ Simba-like & 7 & 5 & 1 & 0 & 2 \\
\ \ Eyeriss-like & 0 & 2 & 0 & 0 & 1 \\
\ \ ShiDianNao-like & 0 & 1 & 0 & 0 & 0 \\
\bottomrule
\end{tabular}
}
\end{table}


Figure~\ref{gra:mohamvssota} shows a subset of Pareto design points obtained by MOHaM along with the design points by CoSA, DiGamma and the best subset of Herald-like for the \textit{AR/VR} workload. Table~\ref{tab:sota_details} presents configuration details of the specific design points ($A$ through $E$) used for comparison. MOHaM's ability to generate Pareto design points offers competitive alternatives to CoSA and DiGamma. For example, design point $A$ from MOHaM reduces latency by $18.40\%$ and energy by $36.87\%$ compared to design point $C$ from CoSA. Despite having a much higher number of MACs, CoSA's latency is higher than MOHaM. It can be attributed to CoSA lacking hardware optimisation and memory-bandwidth awareness. In fact, the whole execution is found to be memory-bounded (refer Table~\ref{tab:sota_details}). The higher number of MACs (therefore PEs) experience increased communication overhead, resulting in higher energy consumption. As CoSA lacks hardware optimisation, its area is considered the minimum configuration of MOHaM to make a fair comparison. Hence, design point $C$ from CoSA has the lowest area among all the design points from Table~\ref{tab:sota_details}.

Similarly, design point $A$ from MOHaM reduces latency by $13\%$ and energy by $96.12\%$ compared to design point $D$ from DiGamma. Despite having a much higher number of MACs (same as CoSA) and hardware-mapping co-optimisation, DiGamma's latency is slightly higher than MOHaM. It can be attributed to DiGamma lacking reconfigurability of the architectural template. For example, DiGamma only has a two-level hierarchical architecture with PE buffers and shared buffers. Whereas MOHaM allows multi-level hierarchy for all the architectural templates supported by the Timeloop cost model. It is worth mentioning that no information about possible memory bottlenecks could be derived from DiGamma output logs. The energy consumption is very high in DiGamma due to large PE buffers. Table~\ref{tab:sota_details} shows that $71.85\%$ of the DiGamma area is occupied by local PE buffers. They generate high-energy reads and writes, resulting in higher energy consumption.

When it comes to Herald-like, design point $E$ is the best EDP solution. Nevertheless, design point $A$ from MOHaM reduces latency by $96\%$ and energy by $93.76\%$ compared to design point $E$. Despite having a hardware-mapping co-optimisation with load-balance driven layer scheduling, Herald-like's latency and energy are much higher than MOHaM. It can be attributed to the random nature of Herald's hardware and mapping search coupled with the unavailability of multi-objective optimisation. Herald tried to improve latency by reducing memory-bound execution. Hence, it allocated large PE buffers and shared buffers, leaving only $5.68\%$ compute area (refer Table~\ref{tab:sota_details}). This action made the execution compute-bound, resulting in higher latency. Moreover, large PE buffers generate high-energy reads and writes, resulting in higher energy consumption. Furthermore, design point $A$ from MOHaM also saves area by $39.78\%$ compared to design point $E$ from Herald-like.

Designers of multi-accelerator systems would benefit from having multiple competitive design points to choose from, especially when there are multiple objectives. However, a mono-objective optimisation aggregates everything into a single design point. This is acceptable as long as there are no conflicting objectives, which is rare. For example, when Herald's mono-objective optimisation aggregated latency and energy together into EDP, it is observed that neither the latency nor the energy is reduced. Herald's attempt to reduce latency by allocating large buffers backfired and resulted into increased latency. Moreover, large buffers are not desirable while attempting energy reduction. Hence, this conflicting scenario resulted in enormous energy consumption. On the other hand, a multi-objective optimisation offers multiple competitive design points while respecting all the objectives and not penalising anyone. For example, MOHaM's multi-objective optimisation allowed for a design point like $A$ that could strike a balance between latency, energy, and area without severely penalising any of them. While there is no \textit{one-size-fits-all}, based on the specific requirements, other design points can also be considered. For example, if energy efficiency is the preference, design point $B$ from MOHaM improves energy by $42.50\%$, $96.47\%$, and $94.32\%$ compared to CoSA, DiGamma, and Herald-like, respectively, at the cost of an increase in latency and area.

\begin{figure}
    \centering
    \includegraphics[width=0.8\columnwidth]{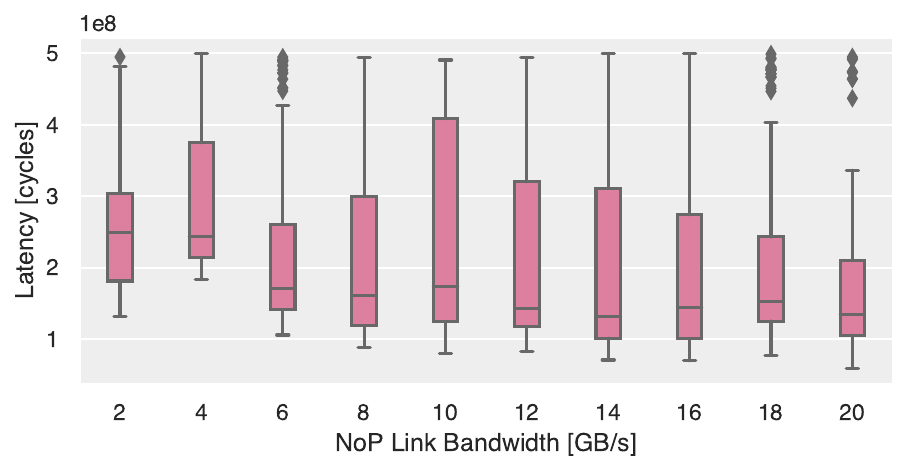}
    \caption{Latency against varying NoP link bandwidth.}
    \label{gra:nop}
\end{figure}

\subsubsection{Sensitivity of NoP Link Bandwidth}
This experiment evaluates the correlation of performance (latency) and NoP link bandwidth. Unlike state-of-the-art, where the communication cost is partially or entirely ignored, MOHaM implements it in detail. The communication model considers head/serialisation delay, routing paths, and bandwidth allocation on NoC and NoP links. Workload latency mainly depends on: (a) \emph{SAI} hosting tile positions in NoP, (b) amount of data accessed from memory as per the layer mappings, and (c) NoP link bandwidth. MOHaM optimises \emph{SAI} hosting tile positions to exploit maximum available bandwidth and avoid bottlenecks. Nevertheless, sometimes it becomes necessary to increase the bandwidth for the desired workload latency. Figure~\ref{gra:nop} shows the latency of Pareto solutions by MOHaM with varying NoP link bandwidth for the \textit{AR/VR} workload. As expected, apart from some exceptions due to the stochastic nature of GAs, there is a trend of decreasing latency with increasing bandwidth. The biggest improvement is observed when the bandwidth increases from 2GB/s to 4GB/s. It is also worth observing that increasing the bandwidth beyond 16GB/s does not guarantee any improvement in latency. This analysis using MOHaM can help reduce communication costs while maintaining the desired performance of the system.

\begin{figure}
\begin{minipage}[c]{0.497\textwidth}
\includegraphics[width=0.9\textwidth]{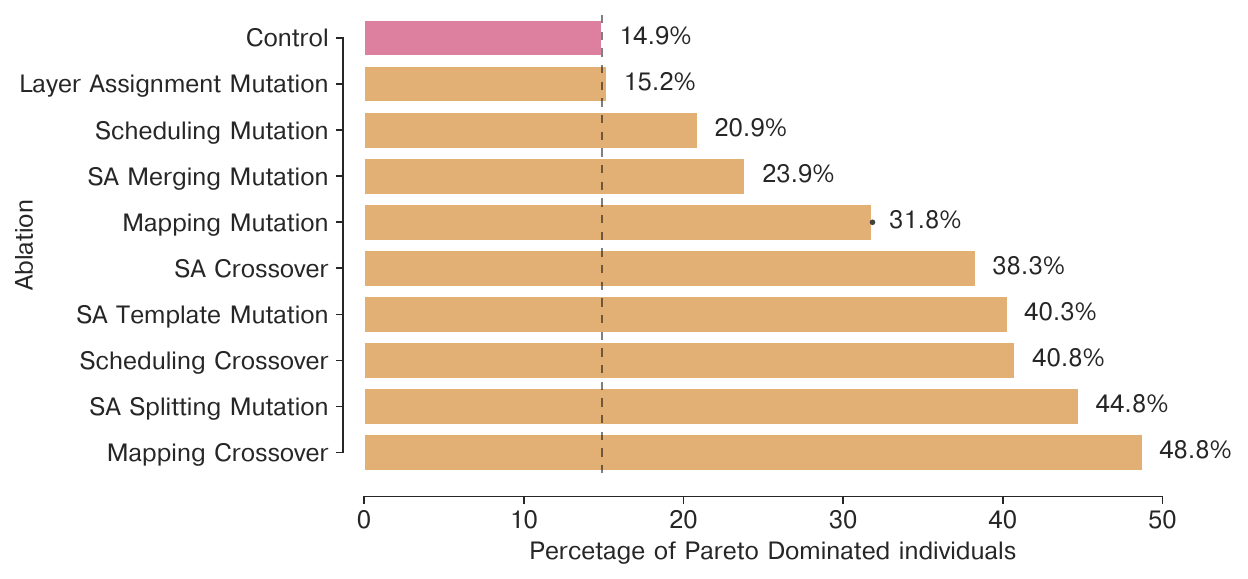}
\caption{Ablation study of the custom GA operators.}
\label{gra:ablation}
\end{minipage}
\hfill
\begin{minipage}[c]{0.497\textwidth}
\includegraphics[width=0.9\textwidth]{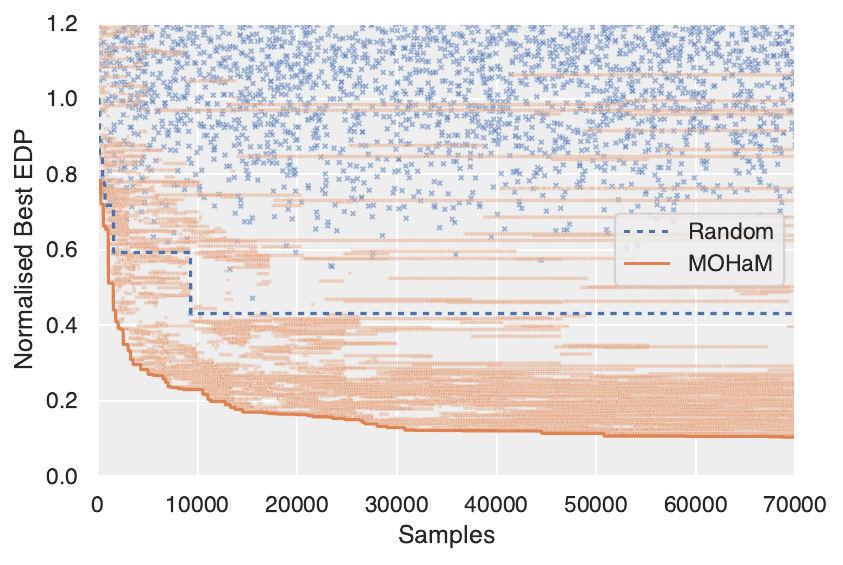}
\caption{Sampling efficiency of MOHaM algorithm.}
\label{gra:sampleefficiency}
\end{minipage}%
\end{figure}

\subsubsection{Ablation Study}
This experiment evaluates the effectiveness of the custom genetic operators shown in Figure~\ref{fig:operators}. It compares the result of a complete MOHaM run with the results of runs, each with one operator disabled (ablated). Figure~\ref{gra:ablation} shows the percentage of Pareto-dominated solutions when an operator is ablated from the baseline MOHaM configuration (refer Table~\ref{tab:mohamconfig}). Due to multi-objective exploration, the results are obtained as follows: \textit{(a)} MOHaM is run with the default configuration, and a baseline Pareto set of individuals is selected from the final population. \textit{(b)} MOHaM is re-run with the same configuration to get a second Pareto set of individuals. \textit{(c)} They are compared to identify how many individuals from the second set are Pareto-dominated by the individuals from the baseline set. This is found to be 14.9\% and serves as the \textit{Control} setting, as shown in Figure~\ref{gra:ablation}. \textit{Control} serves as the threshold to compare ablation results. \textit{(d)} MOHaM is run after disabling one operator to get a new Pareto set of individuals. \textit{(e)} Steps \textit{(c)} and \textit{(d)} are repeated for all the operators, and the results are presented in Figure~\ref{gra:ablation}. A higher percentage of Pareto-dominated individuals indicate that the operator is effective and MOHaM's performance will deteriorate without it. All the operators perform better than the \textit{Control} threshold, implying that each has some significance on MOHaM's performance.

\subsubsection{Sampling Efficiency}
This experiment evaluates the sampling efficiency of the GA-based MOHaM algorithm against a random exhaustive search. Several custom genetic operators are designed for a faster and sample-efficient DSE with MOHaM. Figure~\ref{gra:sampleefficiency} shows the normalised EDP for $70,000$ solution samples for a DSE with random search and a DSE with MOHaM. The trend of solution discovery with the best EDP samples is highlighted with a curve. After these $70,000$ samples, MOHaM achieves $4.17\times$ better EDP compared to the random search. It is observed that some MOHaM solution samples with high EDP continue to be evaluated. This is due to the multi-objective exploration nature of the MOHaM algorithm. While the solution samples from random searches are scattered everywhere, most have poor EDP. The solution curve shows that a DSE with MOHaM will be at least orders of magnitude faster than with a random exhaustive search.

\subsubsection{Execution Time}
In the different experiment settings, MOHaM's execution time did not exceed eight hours, whereas CoSA took only about an hour. CoSA formulates the scheduling space into a MIP problem and solves it in one shot without requiring an iterative search. The scheduling solutions are then evaluated on the Timeloop cost model. Whereas MOHaM formulates the design space into a multi-objective GA problem and solves it by repeatedly querying the Timeloop. While the iterative nature of the GA allows for a more comprehensive exploration of the design space, it also requires more time to converge. Hence, most of MOHaM's execution time is spent repeatedly querying the Timeloop despite attempts to reduce the number of queries through caching. In general, the convergence time of an iterative search-based optimisation is directly proportional to the query response time of the cost model. Therefore, replacing the Timeloop with a faster cost model will ideally improve MOHaM's execution time.

It is important to note that CoSA is a compile-time mapping-only optimisation, so it only makes sense for CoSA to be executed as fast as possible. Whereas MOHaM is a design-time hardware-mapping co-optimisation, so it equally makes sense for MOHaM to be executed without worrying about time for a more comprehensive exploration. MOHaM is used offline and requires a one-time effort. Hence, MOHaM's absolute execution time is of no concern.
\section{Related Works}
\textbf{DNN Models:} Data center workloads are dominated by vision, language and recommendation-based DNN models~\cite{anderson2021first}\cite{richins2020missing}. Vision models are dominated by Convolution (CONV) with some MultiLayer Perceptron (MLP) and Fully Connected (FC) layers towards the end~\cite{he2016deep}. Language models are dominated by MLP, Recurrent Neural Network (RNN), embedding lookup and attention layers~\cite{kenton2019bert}. Whereas, recommendation models mainly consists of MLP, embedding lookup and attention layers~\cite{gupta2020deeprecsys}.
\textbf{Design Space Exploration:} Hardware optimisations are used at design-time by ASICs or even at compile-time by FPGAs. Literature has multiple heuristic as well as ML-based hardware frameworks~\cite{kao2020confuciux}. Mapping optimisations are used at compile-time or even at run-time by reconfigurable accelerators. Literature has mapping frameworks based on heuristics~\cite{yang2020interstellar}, random search~\cite{shao2019simba}, mixed integer programming~\cite{huang2021cosa}, ML~\cite{xiao2021hasco}, etc. Hardware-mapping co-optimisations are used at design-time. Due to the huge cross-coupled search space, very few works have explored co-optimisation for single~\cite{kao2022digamma}\cite{russo2022medea}\cite{xiao2021hasco}\cite{yang2020interstellar} and multi-accelerator systems~\cite{kwon2021heterogeneous}\cite{ghodrati2020planaria}.
\textbf{Multi-Tenancy:} Until recently, multi-tenancy has not been an important design choice for accelerators. Google TPU~\cite{cloud2021tpu}, Microsoft Brainwave~\cite{fowers2018configurable}, etc., focused on running a single DNN model for maximum throughput. The popular MLPerf benchmark suite also focused on a single model for both, training~\cite{mattson2020mlperf} and inference~\cite{reddi2020mlperf}. Data centers are now employing multi-accelerator systems where throughput is not the only objective. Moreover, they have the natural ability to support diverse DNN models as each SA could favour a specific layer with its preferred dataflow. Hence, multi-tenancy has garnered significant attention recently~\cite{kao2022magma}\cite{liu2022veltair}\cite{kwon2021heterogeneous}\cite{ghodrati2020planaria}\cite{choi2020prema}\cite{baek2020multi}.
\section{Conclusion}
This work presents MOHaM, a framework for multi-objective hardware-mapping co-optimisation for multi-DNN workloads on chiplet-based accelerators. The key takeaways are: (1) flexible dataflow with heterogeneous SAs increases the scalability of a multi-accelerator system toward diverse and emerging workloads, (2) hardware-mapping co-optimisation can support diverse workloads with the best overall performance in a multi-accelerator system, (3) multi-objective exploration helps identify the most suitable design for a specific multi-accelerator system, (4) a DSE with the proposed MOHaM framework will be at least orders of magnitude faster than with a random exhaustive search, and (5) replacing Timeloop with a faster cost model will improve MOHaM framework's execution time. A practical scenario where MOHaM could be employed is in designing a DNN-powered Advanced Driver-Assistance System (ADAS), like the ones used in Tesla cars~\cite{tesla}. Another practical scenario could be in designing DNN-based AR/VR hardware for the Metaverse~\cite{meta}. Both of them have known workloads. Future work includes improving the exploration of the scheduling space to enhance the utilisation of the chiplets (SAs).
\section*{Acknowledgments}
This work has been (partially) supported by MUR project ARS01\_00592 reCITY Resilient City Everyday Revolution and by the Spoke 1 FutureHPC \& BigData of the Italian Research Center on High-Performance Computing, Big Data and Quantum Computing (ICSC).
Additionally, author A. Das gratefully acknowledges funding from the European Union’s Horizon Europe research and innovation programme under grant agreement No 101042080 (WINC).

\bibliographystyle{IEEEtran}
\bibliography{references}
\begin{IEEEbiography}[{\includegraphics[width=1in,height=1.25in,clip]{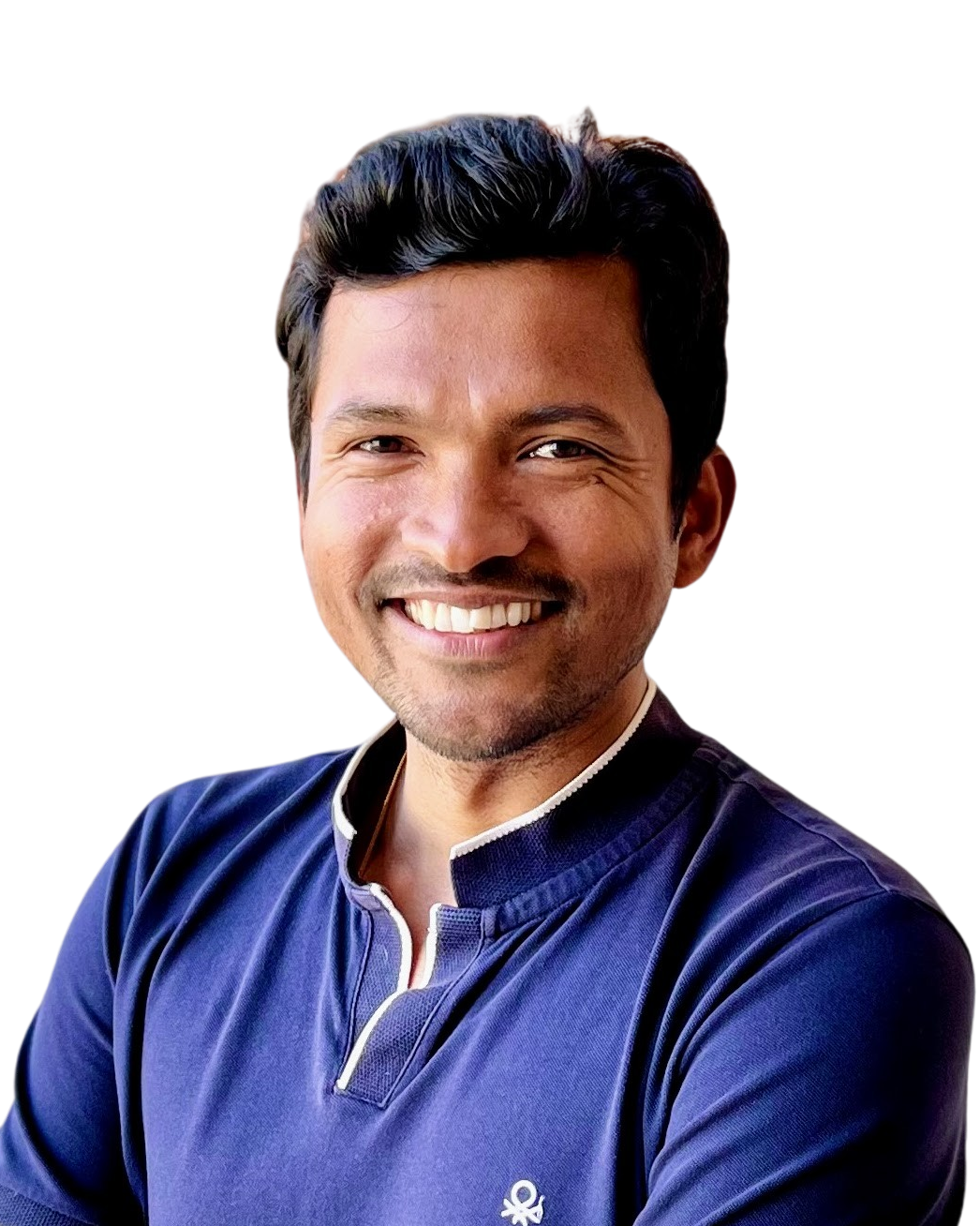}}]{Abhijit Das} is a Post-Doctoral Researcher at the Universitat Politècnica de Catalunya, Spain. He earned the PhD in Computer Science and Engineering from Indian Institute of Technology (IIT) Guwahati, India (2021), and won the Best Thesis Award. His current research interests include chip and package-scale networks, memory systems, DNN accelerators and quantum systems. He is a professional member of IEEE and ACM.
\end{IEEEbiography}
\begin{IEEEbiography}[{\includegraphics[width=1in,height=1.25in,clip]{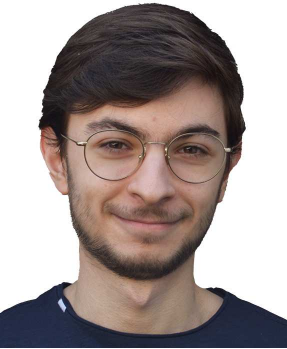}}]{Enrico Russo} is a PhD student at the University of Catania, Italy. Prior to that, he received both his Bachelor's and Master's degrees in computer engineering from the same university. His current research focuses on domain-specific accelerators for Deep Neural Networks (DNNs).
\end{IEEEbiography}
\begin{IEEEbiography}[{\includegraphics[width=1in,height=1.25in,clip]{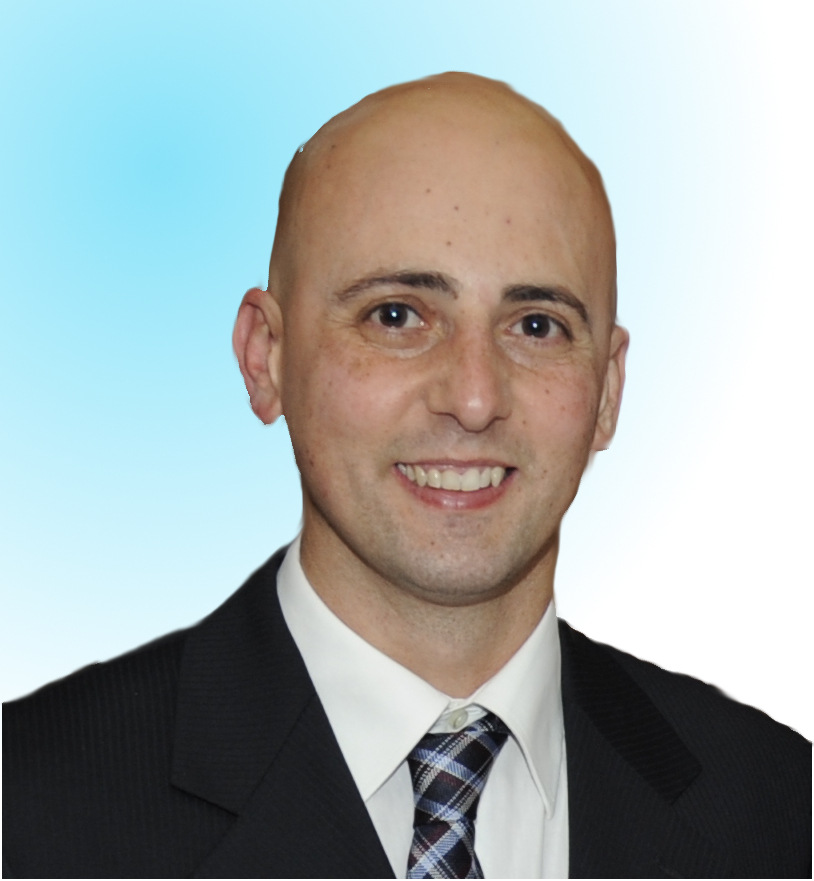}}]{Maurizio Palesi} is an Associate Professor at the University of Catania, Italy, and a Visting Associate Professor at the Indian Institute of Technology Guwahati, India. He received the MSc and PhD degrees in communication and computer engineering from the University of Catania, Italy, in 1999 and 2003, respectively. His current research activity is focused on the area of domain-specific architectures. He was a guest editor for 20 special issues in top-tier journals. He was a general and TPC co-chair for several international conferences and workshops. He is an associate editor for 12 intl. journals. He is an IEEE Senior Member and a HiPEAC Member.
\end{IEEEbiography}

\end{document}